\documentclass{aa}  

\usepackage{graphicx}
\usepackage{txfonts}
\newcommand{\re}{$R_e$}
\newcommand{\Lsig}{$L-\sigma$}
\newcommand{\Lsiga}{$L=L_0\sigma^{\alpha}$}

\newcommand{\Lsigb}{$L=L'_0\sigma^{\beta}$}
\newcommand{\Lsigbtempo}{$L=L'_{0}(t) \sigma^{\beta(t)}$}                   

\newcommand{\MRa}{$R_e$-$M_s$}

\newcommand{\Ie}{$I_e$}

\newcommand{\IeRe}{$I_e-R_e$}
\newcommand{\IeSig}{$I_e - \sigma$}

\newcommand{\FP}{$\log(R_e)= a \log(\sigma) + b \log(\langle I\rangle_e) + c$}

\newcommand{\kmsM}{km\, sec$^{-1}$\, Mpc$^{-1}$}

\def\smallskip{\vskip 6pt}
\def\littleskip{\vskip 2pt}

\begin{document}

   \title{A simple yet effective model of galaxy mergers }


   \author{Cesare Chiosi 
          \inst{1}
          \and
          Mauro D'Onofrio\fnmsep\thanks{Corresponding author Email: mauro.donofrio@unipd.it}
          \inst{1}
          \and
          Emanuela Chiosi
          \inst{2}
          }

   \institute{(1)Department of Physics and Astronomy, University of Padua
              Vicolo Osservatorio 3, I35122, Padua, Italy\\
              (2)PhD in Astronomy, Via Bragni 26D, I35010, Cadoneghe, Padua, Italy \\
              \email{mauro.donofrio@unipd.it, cesare.chiosi@unipd.it, emanuelachiosi@gmail.com}
             }

   \date{Received April, 2026; accepted xxx, 2026}

 
  \abstract
   {In the context of the hierarchical formation of galaxies,  we investigated the role played by mergers in shaping the scale relations  of galaxies, that is the projections of their Fundamental Plane onto  the \IeRe, \IeSig, \MRa\ and \Lsig\ planes. To this aim, we developed a simple model of multiple dry mergers among galaxies by suitably combing the formalism and properties of the so-called infall models of galaxy formation and evolution with the formalism of the  scalar Virial Theorem.  In this context, we  mimicked the hierarchical formation of galaxies and generated simple models of galaxies undergoing a number mergers in the course of their evolution. The results are used  to interpret  the large scale simulations and the companion scale relations  from  observational and theoretical perspectives. }
   {The aim is to interpret  the observational data of the MANGA and WINGS samples and the results of theoretical detailed numerical cosmo-hydro-dynamical simulations, such as  Illustris-TNG100.}
   {In this context, we derived the above scale relations for our theoretical models and compared them  with the observational counterparts from the  MANGA and WINGS database, (and indirectly the large scale simulations of Illustris-TNG100).}
   {The multiple dry merging mechanism  is able to explain all the main characteristics of the observed scale relations of galaxies, such as slopes, scatters, curvatures and zones of exclusion. The distribution of galaxies in these planes is continuously changing across time because of the merging activity and other physical processes, such as star formation,  quenching, energy feedback, and so forth.}
   {The precision of the present simple merger theory is comparable with that obtained by the modern cosmo-hydro-dynamical simulations, with the advantage of providing a rapid exploratory response on the consequences engendered  by different physical effects.}


   \maketitle
%

\section{Introduction}\label{sec:1_intro}

In recent years several studies have addressed the problems posed by the observed scaling relationships (ScRs) of galaxies. It is clear today that the information encrypted in the ScRs is of fundamental importance to understand many physical processes occurring during the evolution of galaxies, such as mergers, stripping, star formation events, feedback, etc.) that in most cases leave imprints on the observed galaxies' distribution in different ScRs. The analysis of the ScRs is therefore of great importance to reconstruct the history of galaxies and the evolution of our Universe.

Many physical parameters of galaxies are mutually correlated, in particular those defining the Virial Theorem (VT) and  the Fundamental Plane (FP, \FP). The mass correlates with the radius (\MRa), the luminosity with the velocity dispersion (\Lsig), the effective surface brightness with the radius (\IeRe),  and others. In log units these relations are often linear, but significant deviations are visible in particular when the distributions of massive and dwarf galaxies are compared. 
In addition, when the sample of galaxies includes objects of very different luminosity, in the ScRs  variations of the mean  slopes  and zones of exclusion (regions empty of galaxies) are visible.

Up to now, most of the efforts of the astronomical community have been dedicated to the analysis of the peculiar characteristics of each ScR. For example, many studies addressed the problem posed by the slope, zero-point, curvature, and scatter.
Other studies highlighted the role played by the environment, feedback effects, dissipation, mergers, Dark Matter halos, galaxy morphology and dynamics, and by cosmology.
A different approach, followed by many researchers, was that of comparing the results of numerical simulations with the observed galaxies in various ScRs. For the sake of brevity, we did not provide references for all the points we touched upon. The reader  can find  exhaustive summaries and proper referencing in   \cite[][]{Donofrio_etal_2024, Donofrio_Chiosi_2024, Donofrio_Chiosi_Brevi_2025}.

The approach we followed  is somewhat  different from the previous ones and finds its roots on the notion that  that  instead of  separately analyzing  the single ScRs, more significant information can be extracted  by simultaneously looking in a self-consistent way at all the mutually related ScRs.
The novelty of this approach is to look at the behavior of each galaxy during its evolution, taking into account  that both luminosity and mass (and consequently all the other related parameters) continuously change for many reasons (star formation, mass acquisition/depletion by mergers or stripping events, natural aging of their stellar populations, and so forth). Method and results are described in a  series of papers
\citep{Donofrioetal2017, Donofrioetal2019, Donofrioetal2020, DonofrioChiosi2021, Donofrio_Chiosi_2022, Donofrio_Chiosi_2023a, Donofrio_Chiosi_2023b,Donofrio_Chiosi_Brevi_2025}. 
to which the reader should refer for more details.

Based on these grounds, \citet{Donofrioetal2017} started from generalizing the luminosity-velocity dispersion relation by explicitly introducing the time dependence in the following way:

\begin{equation}
 L(t)  = L'_0(t) \sigma(t)^{\beta(t)}.
 \label{eq1}
\end{equation}
where $t$ is the time, and $\sigma$ the velocity dispersion. The proportionality coefficient $L'_0$ and the exponent $\beta$ are functions of time and can vary from galaxy to galaxy. Then, by coupling this law with the VT, which is a function of $M_s$, $R_e$ and $\sigma$, they obtained a system of equations in the unknowns $\log L'_0$ and $\beta$ whose coefficients are function of the variables characterizing a galaxy ($M_s$, $R_e$, $L$, $\sigma$, and $I_e$). 
The new empirical relation (\ref{eq1}), although formally equivalent to the Faber-Jackson (FJ) relation for ETGs \citep{FaberJackson1976}, has a profoundly different physical meaning: $\beta$ and $L'_0$ are time-dependent parameters that can vary considerably from galaxy to galaxy and time, according to the mass assembly history and the evolution of the stellar content of each object.
The parameters $\beta$ and $L'_0$ were found to be good indicators of a galaxy's  mass accretion history, star formation, and evolutionary processes, offering an immediate insight into its current stage of evolution. Indeed the $\beta$ parameter determines the direction of motion of a galaxy in the space of parameters at the base of the VT.
Adopting this new perspective, it was possible to simultaneously explain the tilt of the FP and the observed distributions of galaxies in the FP projections and, at the same time, to understand the real nature of the FJ relation. 

Having demonstrated that the ScRs observed today originate from the motion in the parameter space of each individual  galaxy during its evolution, what is still  lacking is to understand which physical phenomena drive the change of shape  of the ScRs in the course of time. In order to reach this goal \citet{Donofrio_Chiosi_Brevi_2025} proceeded  in two parallel ways: 
i) First, they compared the ScRs related to the VT (the $I_e-R_e$, $M_s-R_e$ and $L-\sigma$ relations) with the most recent simulations of model galaxies in cosmological context, such as Illustris-TNG100 \citep{Vogelsberger_2014a, Vogelsberger_2014b}. The comparison was extended to redshift $z\sim 1$ where observations are nowadays available in sufficient number to allow a meaningful analysis, and to even larger  redshift to see what the simulations would predict. 
ii) Then, to cast light on the effects of mergers, in which different masses and energies are involved, on the various ScRs, they set up  a simple analytical model of mergers able to mimic the results of the large scale simulations and to predict the path of galaxies in the parameter space in the course of time. This approach was based on the idea that galaxies in isolation are in virial equilibrium and that a merger is a perturbation of this state of short duration so that a new equilibrium condition is soon recovered. On this ground, the formation and evolution of a galaxy can be conceived as that of an isolated object whose mass and radius time to time are increased by a merger likely causing additional star formation.  
To these aims, \citet{Donofrio_Chiosi_Brevi_2025} generalized the formalism  developed by \citet{Naab_etal_2009} about the material falling onto an already formed object during a single merger, to the case of a series of successive dry mergers among galaxies of different masses. With this simple idea,  \citet{Donofrio_Chiosi_Brevi_2025} were  able to match the observed distribution of galaxies in the ScRs, thus making the dry merging phenomenon the principal mechanism shaping the ScRs observed today. They also showed  that this approach was consistent with the results obtained in previous  studies \citep[see, e.g.][]{Donofrioetal2017, Donofrioetal2020, DonofrioChiosi2021, Donofrio_Chiosi_2022, Donofrio_Chiosi_2023b}. The advantage, with respect to the  sophisticated numerical simulations, resided  in the possibility of obtaining new results on a short time and at no cost, whenever one may want to explore different solutions, such as different initial conditions, different efficiencies of mergers, different laws and modes of star formation, and so forth. 

Although very promising, the merger model developed by \citet{Donofrio_Chiosi_Brevi_2025}  had two points of  weakness. First of all, it needed  an initial set of values for a number of the  physical  quantities characterizing a galaxy (stellar and total mass, effective radius, velocity dispersion, luminosity, etc.) and also a few important parameters of the problem  for which some guess was needed 
\citep[see][for all details]{Donofrio_Chiosi_Brevi_2025}.
Second, important aspects of galaxy structure and evolution were left aside, that is the relative proportions of stars in different age and chemical composition bins, a realistic simulation of the temporal history of mergers and their effects on the galaxies taking part to the event, the temporal history of the mass increase, the luminosity variations, and others.  

In this paper we intend to improve the model of \citet[][]{Donofrio_Chiosi_Brevi_2025} by putting together together galaxies described by the so called ``infall models'' of \citet[][and references therein]{Tantaloetal1998} and a realistic description of a merger  obeying  the virial constraint.

The paper is structured as follow: 
In Sec. \ref{sec:2_data} we presents the observational databases used for our comparisons of the structural parameters with the predictions of theoretical models. 
In Sec. \ref{sec:3_need}  we discuss the need of semi-analytical models of mergers among galaxies. More specifically first we summarize the fully analytical case developed by \citet{Donofrio_Chiosi_Brevi_2025}; second we extensively present the  formalism of semi-analytical model of mergers based on the infall models of galaxies \citep{Chiosi_1980} that is used in this study.
Sec. \ref{sec:4_resana}  contains a detailed description of the results obtained from our analytical models and a comparison with the observational data at $z\sim0$. 
In Sec. \ref{sec:5_betatheory} we discuss the results at the light of the $\beta-L_0'$ theory of \citet[][and references therein]{Donofrio_Chiosi_2024} and explain the reason for the wide dispersion in $\beta$ shown by the observational data. 
In Sec. \ref{sec:6_comp} we compare our models with those of the Illustris-TNG100 simulations at different values of the redshift and outline some interesting differences with respect to the case at $z=0$.
Finally, in Sec. \ref{sec:10_concl} we draw some conclusions.  The paper, contains also three Appendices A, B, C in which we summarize ideas, models, and results already developed in previous studies but that are needed here to help the reader understand the present paper.

\section{Observational data, modern large scale simulations, and adopted cosmological model of the Universe}\label{sec:2_data}

In this study we make use of two samples of observational data:  

(i) The primary galaxy sample is that of MANGA (Mapping Nearby Galaxies at APO) survey \citep{Smee_etal_2013, Bundy_etal_2015, Drory_etal_2015, Blanton_etal_2017}. MANGA obtained spectral and photometric measurements of nearby galaxies of different morphological types ($\sim10000$ objects across $\sim2700$ sky square degree), thanks to the “integral field units” (IFUs) of the Apache Point Observatory (APO). MANGA is a byproduct of the SDSS survey and was designed to study the history of present day galaxies.
MANGA provided a lot of galaxy parameters: morphology, stellar rotation velocity and velocity dispersion, magnitudes and radii, mean stellar ages and star formation histories, stellar metallicity, element abundance ratios, stellar mass surface density, ionized gas velocity, star formation rate, dust extinction, and others. The galaxies were selected to span a stellar mass interval of nearly three orders of magnitude and no selection was made on size, inclination, morphology and environment, so this sample is fully representative of the local galaxy population and can be compared with the modern hydro-dynamical simulations like Illustris-TNG, which, however, do not provide information on the morphological types of their model galaxies.
The present sample contains information for 10220 galaxies of all morphological types.

(ii) In parallel, we also took into account the WINGS and Omega-WINGS databases
\citep{Fasanoetal2006,Varela2009,Cava2009,Valentinuzzi2009,Moretti2014,Donofrio2014,Gullieuszik2015,Morettietal2017,Cariddietal2018,Bivianoetal2017}, already adopted in our previous studies on the subject \citep[see,][]{Donofrio_Chiosi_2022,Donofrio_Chiosi_2023a}. The WINGS photometric sample contains about 33000 galaxies. In addition we set up a smaller sub-sample  containing  only 270 early-type galaxies (ETGs), generally with stellar mass $M_s > 10^9 \, M_\odot$; the objects in this sub-sample have  measurements of their stellar masses, star formation rates (measured by \citet{Fritzetal2007}), morphologies (given by MORPHOT \citep{Fasanoetal2012}), velocity dispersions, effective radii,  and total luminosities. The two samples contain objects with  z$\sim 0$.  We will make use of both samples depending on circumstances.

(ii) Finally, we extended our observational database making use of the  classical catalog of nearby galaxies by \citet{Burstein_etal_1997} which includes data from  globular clusters  to  galaxy clusters. Although this source of data may be considered as nowadays superseded by the more recent surveys we have already quoted above, we consider it as still worth of interest also in  modern studies of the ScRs.

(iii) The large simulations of galaxy formation and evolution in the hierarchical scheme we have adopted here for comparison with our simple analytical model are those of  Illustris-TNG \citep{Springeletal2018,Nelsonetal2018,Pillepichetal2018b}. The new simulation seems to produce results in better agreement with the observations \citep{Nelsonetal2018, Rodriguez-Gomezetal2019}. In Illustris-TNG the structural parameters of galaxies, such as colors, magnitudes and radii, are in quite good agreement with the available data at low redshift \citep{Rodriguez-Gomezetal2019}. Here we used the Illustris-TNG100 dataset. 
For the Illustris-TNG100 simulations, the main progenitor branches of the merger trees were considered to follow the evolution of the same objects along time. The more massive galaxy has been always selected as the progenitor of each object going back to high redshift, from z=0 to z=4, in steps of 1.
Only those object that were already formed at z=4, with a stellar mass higher than the mass resolution of the simulations ($M_{res} = 1.4\times 10^6 M_\odot$) at all redshift, and have a stellar mass $M_s > 10^{8.5}$ at z=0,  were considered.
To have data sets that are easier to handle and at the same time to keep the statistical robustness, only a maximum of 150 objects for each 0.2 dex of stellar mass at z=0 were considered.
At the end, a total of 2200 objects were taken into consideration in Illustris-TNG100.

We anticipate here that throughout the paper the present theoretical results are extensively compared to the observational data whereas large scale simulations are left in background.  These later have already been compared to each other and to observational data by the authors themselves and in other papers of ours of this series \citep[see for instance][and references]{Donofrio_etal_2024,Donofrio_Chiosi_2024,Donofrio_Chiosi_Brevi_2025}. In general there is satisfactory agreement among the various sources of data and theoretical models.

Before closing this section,  we recall that we frame the present analysis  in the $\Lambda-CDM$ cosmology whose parameters are $\Omega_m$ = 0.2726, $\Omega_{\Lambda}$= 0.7274, $\Omega_b$ = 0.0456, $\sigma_8$ = 0.809, $n_s$ = 0.963, $H_0 = 70.4\,km\, s^{-1}\, Mpc^{-1}$, the same values adopted in 
previous studies of this series for. This choice allow us to compare our results  with the data of the Illustris-1 large scale hierarchical galaxy simulations of   \citet{Vogelsberger_2014a,Vogelsberger_2014b}.
Slightly different cosmological  parameters are used by for the Illustris-TNG simulations: $\Omega_m$ = 0.3089, $\Omega_{\Lambda}$= 0.6911, $\Omega_b$ = 0.0486, $\sigma_8$ = 0.816, $n_s$ = 0.967, $H_0 = 67.74\,km\, s^{-1}\, Mpc^{-1}$ \citep{Springeletal2018,Nelsonetal2018,Pillepichetal2018a}. However, since the systematic differences  in $M_s$, $R_e$, $L$, $I_e$, and $\sigma$ are either small or nearly irrelevant to the aims of this study, no re-scaling of the data is applied.

 \section{ Need of semi-analytical models of galaxy mergers} \label{sec:3_need}
  
The theoretical large scale simulations we referred to  in the previous section were  calculated in the hierarchical scenario in which mergers among galaxies are the current paradigmatic view of galaxy  formation and evolution. Since data and theory were in satisfactory agreement, the conclusion was that the hierarchical view was the right frame to work with and to highlight the role played by the merger mechanism in shaping the various scale relations followed by galaxies.  However, the large scale simulations are so heavy to plan, perform, and interpret that changing the physical ingredients to explore their effects  on the ScRs of galaxies is a cumbersome affair. Therefore, in \citet{Donofrio_Chiosi_Brevi_2025} and in this study we developed a simple analytical model of mergers as a smart proxy of the large scale numerical simulations and considered  the task as a meaningful and hopefully useful effort.

\subsection{The fully  analytical merger model}\label{subsec:3a}
    
The \citet{Donofrio_Chiosi_Brevi_2025} model strictly stands on the formalism  developed 
by \citet{Naab_etal_2009} for the case of a single object in virial equilibrium on which another object of assigned total mass and energy falls in and merges. In a few words, let us consider a galaxy characterized by the parameters  $M_i$, $Re_i$,  $\sigma_i$, $L_i$, and $Ie_i$  before the merger and obeying the virial conditions  expressed by  the relations among the kinetic ($K_i$),  gravitational ($W_i$),  and total ($E_i$)  
 
 \begin{eqnarray}
 K_i = \frac{1}{2} M_i \sigma_i^2, \quad  W_i = - \frac{G M_i ^2}{ R_{e,i}},  \quad  
 E_i = K_i + W_i;
 \end{eqnarray}
 \begin{eqnarray}
 K_i = - \frac{W_i}{2};  \quad   E_i =  \frac{W_i}{2}.
 \end{eqnarray}

 \noindent
 Suppose now that the galaxy {\it i}  merges  another galaxy {\it a} with  parameters $M_a$, $R_{a,i}$, $\sigma_a$, $L_a$,  $I_{a,i}$, $K_a$,  $W_a$,  and $E_a$, and that after merging the virial conditions are rapidly restored. We can express the mass and the velocity dispersion of the accreting galaxy by means of the same quantities for the initial object, that is by  the parameters  $\eta= M_a/M_i$ and $\epsilon = \sigma_a^2 /\sigma_i^2$. The total energy of the system is $E_f = E_i + E_a$. After merging, the composite system recovers the virial equilibrium so that we can write
 
 \begin{equation}
  E_f = - \frac{1}{2} M_f \sigma_f^2  = - \frac{1}{2} M_i \sigma_i^2 (1 + \epsilon \eta)  
 \end{equation}

 \noindent
 where $M_f = M_i + M_a = (1+ \eta) M_i$.  Finally, we obtain the following basic relationships 
 
 \begin{equation} 
  \frac{M_f}{M_i}   = (1+ \eta);  \quad\quad
  \frac{\sigma_f^2}{\sigma_i^2}  = \frac{1+ \eta \epsilon}{1+\eta};  \quad\quad
  \frac{R_{e,f}}{R_{e,i}}        = \frac{(1+\eta)^2}{1+ \epsilon \eta}. 
 \end{equation}

\noindent 
If required, one can also  add a relationship between the final $L_f$ and the initial $L_i$ luminosities. 
More details on the formalism and equations developed by \citet[][]{Donofrio_Chiosi_Brevi_2025}  are given in Appendix \ref{app1_ana_merger}.

For a more detailed description  see \citet[][]{Donofrio_Chiosi_Brevi_2025} who  generalized the whole problem and formalism to the case of  many successive mergers with (i) the same mass and dispersion velocity; (ii) different mass and velocity dispersion; (iii) transfer of  a fraction of the kinetic energy of the relative motion of the two galaxies to the composite system  
of the captured galaxy in addition to the energy associated to the stellar velocity dispersion; (iv) effects of random variations from merger to merger of the parameters defining a merger; (v) finally, updating the  reference model  after each merger, that is  the mass $M_f$ and the velocity dispersion $\sigma_f$ resulting from a merger event are the initial value for the next one.

The results obtained using the above simple analytical description of mergers and their comparison with observational data for the scaling relations in several diagnostic planes were very promising thus spurred us to proceed further along this line of thought.

There were two points of weakness in the above simple model. First of all, it needed  an initial set of values for the  physical quantities $M_i$, $\sigma_i$,  $R_{e,i}$, $L_i$, $I_{e,i}$ together with some guess for the parameters $\eta$, $\epsilon$, $\theta$ related to luminosity and $\lambda$ (related to energy sharing)  for these latter see Appendix \ref{app1_ana_merger}. 
Second, important aspects of galaxy structure and evolution were left aside, that is the relative proportions of stars in different age and chemical composition bins, a realistic simulation the temporal history of mergers and their effects on the galaxies taking part to the event, the temporal history of the mass increase, the luminosity variations, and others.  

In this paper we intend to improve the model of \citet[][]{Donofrio_Chiosi_Brevi_2025} by merging together galaxies described by the so called ``infall models'' of \citet[][and references therein]{Tantaloetal1998} and obeying the virial conditions represented by the equations

 \begin{eqnarray}
  \frac{M_f}{M_i}  &=& (1 +  \eta) \nonumber \\
  \frac{\sigma_f^2}{\sigma_i^2} &=& \frac{ (1 + \eta (\epsilon+\lambda) )}{(1+  \eta)} \nonumber \\
  \frac{R_{e,f}}{R_{e,i}}       &=& \frac{(1 + \eta)^2}{ (1 +  \eta (\epsilon + \lambda) )}  \\
  \frac{L_f} {L_i}              &=& (1  +  \eta \theta)   \nonumber \\
  \frac{I_{e,f} }{I_{e,i}}      &=& \frac{ (1 + \eta \theta) (1+ \eta (\epsilon+\lambda))^2 }{(1+  \eta)^4} \nonumber   
  \end{eqnarray}
  \label{Var_Ref_Gal_a}

  \noindent
derived by \citet[][]{Donofrio_Chiosi_Brevi_2025}  for the case (v) listed in Appendix \ref{app1_ana_merger},  in which at each merger the masses $M_i$ and $M_f$ and the luminosities $L_i$ and $L_f$  are known from the galaxies undergoing the merger,   and parameters $\eta$, $\epsilon$,  $\lambda$,  and $\theta$ are suitably recast from the available information about the two merging objects.

The new modeling of a merger stands on an important property of the ``infall models'' of galaxy formation and evolution, that is the self-similarity of their evolutionary history at varying the total galaxy mass while keeping constant the time scale of mass accumulation. To clarify the issue and help the reader of our paper, in Appendix \ref{app2_open_mod} we shortly review the basic assumptions of the infall models and outline the main features of their evolution.

  \begin{figure}        
   \centering
   \includegraphics[scale=0.25]{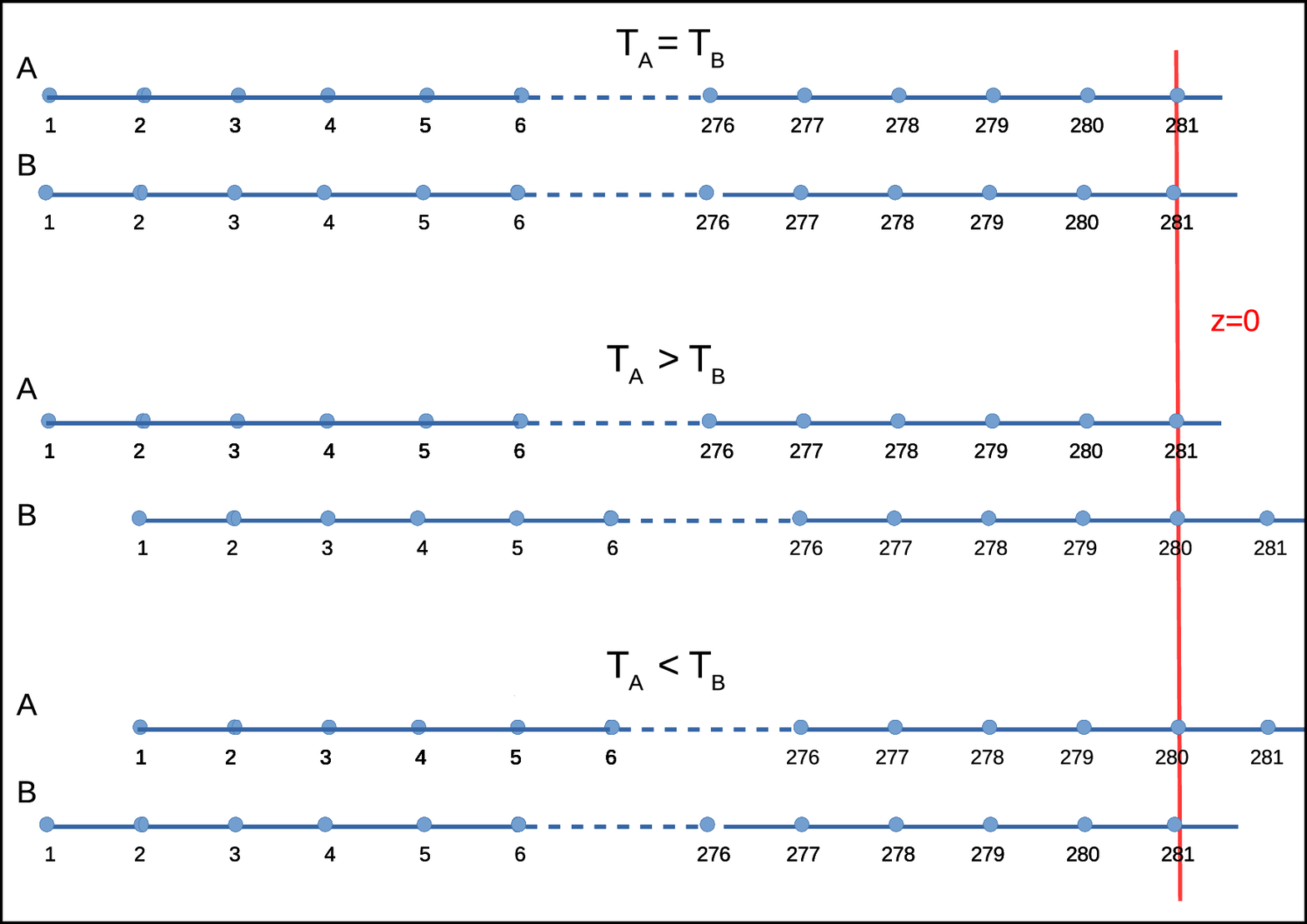}
\caption{Slide rule mimicking a merger. The cases on display correspond to $\Delta_{AB}=0$ ($T_A = T_B$), $\Delta_{AB}=1$ ($T_A \leq T_B$) and 
   $\Delta_{BA} =1$ ($T_A \geq T_B$). The vertical red bar marks the redshift $z=0$ or equivalently the age $T_G=13.187$ Gyr.}
              \label{Fig:ODG}
    \end{figure}

\subsection{The semi-analytical merger history of a galaxy} \label{subsec:3b}
A galaxy model with infall is characterized by a number of physical quantities such as the current baryonic  mass $M_b(t)$, the stellar  mass $M_s(t)$, the gas mass $M_g(t)$,  the abundances (by mass) of elements $X_j(t)$, the metallicity $Z(t)$ (sum of all elements heavier than He), the mean metallicity $<$Z(t)$>$, the infall rate of baryonic mass $dM_b(t)/dt$, the star formation rate $R_s(t)= - \nu dM_g(t)/dt =  dM_s(t)/dt$ (where $\nu$ is the specific efficiency of star formation), the luminosity $L(t)$ (and luminosities and/or magnitudes in a suitable photometric system), together with the half-mass radius $R_e(t)$, and the mean velocity dispersion $\sigma(t)$, the specific intensity $I_e(t)$, and finally the age. Each quantity is expressed in typical suitable units; masses in $M_\odot$, star formation and infall rates in $M_\odot / Gyr$, luminosities in $L_\odot$, radii in kpc or pc, velocity dispersion in km/s, and ages in Gyr. $t_G$ is the present age of the galaxy in Gyrs. Each model is labeled by its final baryonic mass  $M_B(t_G)$ expressed in solar units. 

For a number of them one can define the corresponding dimensionless variables, chief among others are the fractionary  gas mass $G_g(t)= M_g(t)/M_b(t_G)$ and  stellar mass $G_s(t)= M_s(t)/M_b(t_G)$.

At given time scale of baryon mass accretion, $\tau$, and star formation efficiency $\nu$, by construction, the time dependence of the dimensionless quantities is always the same regardless of the total mass. Similar behavior applies to the time dependence of the total luminosity per unit mass 
$L(t)/M_b(T_G)$ and companion luminosities per unit mass $l_{\Delta\lambda}(t)$ in the 
$\gamma_{th}$ pass-band  of the adopted photometric system, shortly indicated as $\Delta\lambda_\gamma$. 

This is a net advantage offered by infall models, because once the time scale $\tau$, the specific star formation efficiency $\nu$, and the stellar initial  mass function (IMF),  are assigned, a manifold of models is soon available each of these characterized by the value  of the baryon mass $M_b(t_G)$, therein after simply refereed to as the \textit{asymptotic baryon galaxy mass} or shortly the \textit{baryon galaxy mass}.

 Therefore when two galaxies of baryon mass $[M_{b}(t)]_A$ and $[M_{b}(T)]_B$, shortly indicated as $\rm M_A$ and $\rm M_B$, merge together to form the composite galaxy with  mass $M_C = M_A + M_B$,  the gas content, star contents, star formation rate,  and metallicity are given by the relations: 
 \littleskip
$M_{g,C}(t)= M_{g,A}(t)+M_{g,B}(t)$,  

$M_{s,C}(t)= M_{s,A}(t)+M_{s,B}(t)$ , 

$  {R_{s,C}(t)} = (  {R_{s,A}(t)}  +  {R_{g,B}(t)} ) $,  
\littleskip
\littleskip
\noindent
 while the current metallicity $Z(t)$ and current mean metallicity $<Z(t)>$ can be summed up each one weighed on the ratio $M_A /M_C$ or $M_B/M_C$ as appropriate
\littleskip
 $ Z(t) = (Z_A(t) M_A +Z_B(t) M_B) /M_C(t)$ 
\littleskip
 \noindent
and 
\littleskip
$ <Z(t)> = ( <Z_A(t)> M_A + <Z_B(t)> M_B) /M_C(t)$
\littleskip
\noindent
finally for the current total mass 
\littleskip
$M_{C} = M_A + M_B$.
\littleskip
\noindent
The same for the luminosities  $l_{\Delta\lambda}(t)$ of the composed galaxies which are given by
\littleskip
$L_{\Delta\lambda,C}(t) =  l_{\Delta\lambda,A}(t) * M_A + l_{\Delta\lambda,B}(t) *M_B $.
\littleskip
\noindent
As a consequence of a merger, the contents of stars, gas, total mass, luminosity, etc. change and must be redefined soon after the merger is over. 
 
It is convenient for each physical quantity of interest to store the change occurred at each value of the age. In  other words at each time step of the history of a galaxy we store the changes occurred in the above quantities for each value of the age, that is past present (and future) so that  all information necessary to reconstruct the real  path followed by a galaxy during its merger history  is saved. To clarify this important issue suffice to recall that at each time of a galaxy's life, all  integrated properties, such as for instance the total light, are  the integral over the past of the contribution of the sub-populations of different age and chemical compositions present in the mix in different proportions (relative mass and/or relative numbers) due to  the individual  history of each galaxy taking part to the merger event and also the new populations added by star formation at the time of the merger.

A generic merger  is described as follows: 

(i) We start from  the infall models of galaxies whose evolutionary history if calculated from 
$t_G$ = 0 at the formation redshift $z_f$  to the present galaxy age $t_G$ at $z=0$, the age $t_G$ is referred to as the rest-frame age of a galaxy. The  age $t_G$ 
depends of the cosmological model of the Universe and the formation redshift $z_f$. In  the $\Lambda$-CDM Universe with  parameters $\Omega_m$ = 0.2726 (of which $\Omega_b$ = 0.0456 in baryons), $\Omega_{\Lambda}$= 0.7274,   $H_0 = 70.4\,km\, s^{-1}\, Mpc^{-1}$   and redshift of first galaxy formation $z_f$ = 10 we have adopted,   we have age of Universe $T_{u,z_f} = 0.484$ Gyr, present age of the Universe   $T_{u,0} = 13.671$ Gyr or equivalently rest-frame age of the first galaxies  $t_G=13.187$ Gyr. Other choices for $z_f$ and the cosmological parameters are possible. 
We also assume the time scale of mass accretion $\tau = 1$ Gyr and specific efficiency of the star formation rate $\nu =1$. Also in this case, other choices are possible. 
The models are calculated with good accuracy so that epochs of fast and slow mass accretion by infall and star formation are followed with care. The whole history of a galaxy is described as a function of an age vector  containing 281 values of the rest-frame age to which an equal number of $T_{u,z}$ and redshift $z$ are associated.  For each model we know all the variables $Y_i(t)$ (where $i$ stands for stars, gas,  etc.). The age-redshift grid is considered fixed in our description of a  merger. 
Finally, we distinguish between the galaxy, named A, capturing merging galaxies, and the galaxies, named B, that merge into the  galaxy A.  The galaxy A is supposed to form at $z_f=10$, whereas galaxies B  may  form at any redshift and hence have any age from $t_G=0$ to $t_G$= 13.187 Gyr. 

(ii) The age vector or grid is considered as a sort of  slide ruler along which 281  notches are marked, each one identified by the number $N$  and the rest-frame age $t_G$. Given two galaxies of mass $M_A$ and $M_B$, age  $t_A$  and $t_B$ that are going to merge at some age $T_{u,m}$ or rest-frame $t_m$, the slide rule must be set up according to the situation we intend to  model. First of all, the mass of each object and its age are randomly chosen within some reasonable intervals, second we must establish the mutual relationship between each pair of variables of the same type. The masses can be equal or different, that is  $M_A = M_B$, $M_A > M_B$, and $M_A < M_B$ are possibile. As far the ages are concerned, we have $t_A > t_B$, $t_A = t_B$, 
and $t_A < t_B$. This latter possibility, however,  is suject to some restrictions.  

(iii) A few short  comments on the ages of galaxies A and B are appropriate. Ideally, the situation should be as follows: select $z_f$ (lower than 10) for the galaxy A and calculate the associated infall model A, selected the  series of ages $t_{A,m}$ at which mergers with this galaxy are expected to occur, select masses and ages for the the merging galaxies B, no severe limits on the mass, and formation reshift and ages obeing the constraint $z_f < 10$. This ideal case is not yet possible because the available infall models are all calculated with $z_f$=10.  
Since galaxy A has been formed at $z_f$ =10 while galaxies B can be formed at any  $z_f$ in the interval 10 to 0, this implies that while there is  no constrain on $t_A$, a obvious constraint exists  for $t_B$. Therefore, the case $t_B = t_A$ is very unlikely but possible, the case $t_B < t_A$ is likely and possible,  the case $t_B > t_A$ must be rejected because  $t_B$ cannot be older than $t_A$. If randomly selecting the mass  and age  for the B galaxy (see below) this constraint is violated, the choice is rejected and other values are chosen. 
How severe is the restriction imposed by the current infall models on the ideal situation? Fortunately, thanks to the short infall time scale $\tau=1$ Gyr we have adopted, a way out is possible. With such short time scales, all the action (mass growth,  peak of the star formation rate, increase of the stellar content, decrease of the gas content, increase of metallicity, and so forth) takes place within  the first 1.5 to 2 Gyr and  then all these quantities level off or even gently decline. Consequently, if we take a
one of the B galaxies  at a certain age sufficiently   older than 1.5 to 2 Gyr or so, to a good approximation we can consider as if it were born 2 to 3  Gyr earlier than the current age. This is equivalent to say that at any age galaxies B can be considered as a reasonable approximation of objects born in a recent past. We will make use of the this approximation. In any case, work is in progress to remove this limitation of the ages of the galaxies B.    
Finally, the times $t_A$,  $t_B$, and  $t_m$  find direct correspondence in the age-redshift-grid with the age-redshift bin numbers $N_A$, $N_B$ and $N_m$.

iii) From an operational point of view it is convenient to proceed as follows. We randomly choose the total number of mergers ($N_k$) we want to consider,  randomly choose $M_A$ and  the age $t_A$ at which the first merger is supposed to occur, randomly choose $M_B$ and $t_B$, however keeping in mind the constraints we examined above, the random choice is repeated in the case of failure.  Having set the mass and age of the first merger, then we determine the parameters for all the merger sequence up to $N_k$, that is we select the series of ages $t_A$ for the galaxy A at which subsequent mergers should occur and select series of the masses  $M_B$  and ages $t_B$ of the galaxy B involved in the series of mergers. At each step, the $M_A$ is increased by the mass $M_B$  and the new mass $M_A$ is updated to $M_A + M_B$.   Finally, all ages $t_A$ and $t_B$ are rounded to the nearest value in the age-redshift-grid. By doing so, the age (and redshift) can be replaced by the grid numbers J's.
The net advantage is  that all physical quantities of interest can be grouped in $K,J, V$ tables where  $k=1, ....N_m$ (merger number),  $J=1,...281$ (age number) and $V=i,...V_m$ (index of  the variable of interest in the list of physical quantities describing the status of a galaxy). The dimensions of the tables  and number of  indices can be varied (i.e. either increased or decreased) as needed. The retrieval of any quantity of interest is fast and safe.

iv) In order to update the generic quantity  (gas, stars, star formation rate, metallicities, current mass)  to the new situation after each merger, we proceed as follows. At the time of merger the slide rules representing each galaxy must  shift back or forth so that the notches $N_A$ and $N_B$ are at the same position. The situation is illustrated in Fig. \ref{Fig:ODG} for the three cases $t_A=t_B$, $t_A > t_B$, and $t_A < t_B$. Although, this case never happens in this study because on the age limits on 
$t_B$ we still describe it in view of future applications. The vertical red line correspond $j=281$, that is time $T_G$ or redshift $z=0$, the present. If the slide rule of a galaxy goes beyond this boundary, all points (age bins) at the right side must neglected. In the same way, the galaxy can have the  age bins at the left of the first age bin of the other galaxy. This means that in this galaxy all those age bins are not affected by the merger. 

\noindent
Introducing the difference 
$\Delta_{AB} = N_A - N_B$  or $\Delta_{BA} = N_B -N_A$ as appropriate, we have the following relationships:

{\bf Case (A)} - The merging galaxies have the same age, therefore
  $t_A = t_B,  \quad  N_A = N_B$ \quad $\Delta_{AB} = N_A - N_B = 0$.

\begin{eqnarray} 
M_{C}        &=& M_A + M_B  \nonumber \\
M_{g,C}(j)   &=& M_{g,A}(j)+M_{g,B}(j)  \nonumber \\  
M_{s,C}(j)   &=& M_{s,A}(j)+M_{s,B}(j) \nonumber \\ 
{R_{s,C}(j)} &=&   {R_{s,A}(j)}  +  {R_{s,B}(j)}   \\  
  Z(j)       &=& (Z_A(j) M_A +Z_B(j) M_B) /M_C \nonumber \\ 
 <Z(j)>      &=& ( <Z_A(j)> M_A + <Z_B(j)> M_B) /M_C \nonumber \\
  L_{\Delta\lambda_{\gamma},C}(j) &=&  l_{\Delta\lambda_{\gamma},A}(j)  M_A + l_{\Delta\lambda_{\gamma},B}(j) M_B  \nonumber
\end{eqnarray}

\noindent   
for $j=1,....,281; \quad  k=1,...., N_k $, \quad $\gamma=1,...n_{\Delta\lambda}$; \\

{\bf Case (B)} - The galaxy A is older than galaxy B, therefore we have  $t_A > t_B$,   $N_A > N_B$, and $\Delta_{AB} = N_A - N_B$.  The evolution of the receiving galaxy splits in two parts, a first one in which the A galaxy evolve alone, and a second one after the merge in which its properties are modified by the effect of engulfing the galaxy. In terms of the $j$ index describing the age we have:

\noindent
B1) When $j$ is lower than or equal to  $\Delta_{AB}$, the following equations apply

\begin{eqnarray} 
M_{C} &=& M_A  \nonumber \\
M_{g,C}(j) &=& M_{g,A}(j) \nonumber \\  
M_{s,C}(j) &=& M_{s,A}(j)   \nonumber \\ 
{R_{s,C}(j)} &=&   {R_{s,A}(j)}    \\  
  Z(j) &=& Z_A(j)  \nonumber \\ 
 <Z(j)> &=&  <Z_A(j)>   \nonumber  \\
  L_{\Delta\lambda_{\gamma},C}(j) &=&  l_{\Delta\lambda_{\gamma},A}(j) M_A  \nonumber
\end{eqnarray}

\noindent
B2) When $j$ becomes greater than $\Delta_{AB}$, we must define a new index for the galaxy B given by 
$j_B = j + \Delta_{AB}$ and use the equations

\begin{eqnarray} 
M_{C} &=& M_A + M_B \nonumber  \\
M_{g,C}(j) &=& M_{g,A}(j)+M_{g,B}(j_B) \nonumber \\  
M_{s,C}(j) &=& M_{s,A}(j)+M_{s,B}(j_B) \nonumber  \\ 
{R_{s,C}(j)} &=&  {R_{s,A}(j)}  +  {R_{s,B}(j_B)}   \\  
  Z(j) &=& (Z_A(j) M_A +Z_B(j_B) M_B) /M_C \nonumber  \\ 
 <Z(j)> &=& ( <Z_A(j)> M_A + <Z_B(J_B)> M_B) /M_C   \nonumber  \\
  L_{\Delta\lambda_{\gamma},C}(j) &=&  l_{\Delta\lambda_{\gamma},A}(j)  M_A + l_{\Delta\lambda_{\gamma},B}(j_B) M_B \nonumber
\end{eqnarray}
  
\noindent
  for   $j=1,...., 281$;  $k=1,...., n_k $, and $\gamma=1,...n_{\Delta\lambda}$; \\

{\bf Case (C)} - The galaxy A is younger than galaxy B, therefore we have $t_A < t_B,   N_A < N_B$,  and $\Delta_{AB} = N_B - N_A$.  The situation is similar to that of Case (b) but now the galaxies exchange the role. The evolution of galaxy B splits in two parts. In the first one,  the galaxy B evolve alone, the second one the two galaxies merge and evolve together. The equations governing Case (c) are

\noindent
C1) When $j$ is less or equal to  $\Delta_{BA}$, the following equations apply

\begin{eqnarray} 
M_{C} &=& M_B  \nonumber   \\
M_{g,C}(j) &=& M_{g,B}(j)   \nonumber  \\  
M_{s,C}(j) &=& M_{s,B}(j)   \nonumber \\ 
{R_{s,C}(j)} &=&   {R_{s,B}(j)}    \\  
  Z(j) &=& Z_B(j)   \nonumber \\ 
 <Z(j)> &=&  <Z_B(j)>   \nonumber  \\
  L_{\Delta\lambda_{\gamma},C}(j) &=&  l_{\Delta\lambda_{\gamma},B}(j)  M_B  \nonumber 
\end{eqnarray}

\noindent
C2) When $j$ becomes greater $\Delta_{BA}$ we must define a new index for the galaxy A given by 
$j_A = j - \Delta_{BA}$ and use the equations

\begin{eqnarray} 
M_{C} &=& M_A + M_B    \nonumber \\
M_{g,C}(j) &=& M_{g,A}(j_A)+M_{g,B}(j)  \nonumber  \\  
M_{s,C}(j) &=& M_{s,A}(j_A)+M_{s,B}(j) \nonumber  \\ 
{R_{s,C}(j)} &=&  {R_{s,A}(j_A)}  +  {R_{s,B}(j)}   \\  
  Z_C(j) &=& (Z_A(j_A) M_A +Z_B(j) M_B) /M_C   \nonumber  \\ 
 <Z_C(j)> &=& ( <Z_A(j_A)> M_A + <Z_B(J)> M_B) /M_C \nonumber  \\
  L_{\Delta\lambda_{\gamma},C}(j) &=&  l_{\Delta\lambda_{\gamma},A}(j_A)  M_A + l_{\Delta\lambda_{\gamma},B}(j) M_B \nonumber
\end{eqnarray}
  
\noindent
for   $j=1,...., 281$;  $k=1,...., N_k $, and $\gamma=1,...n_{\Delta\lambda}$; \\

v) Before moving to the next merger, all variables of the mass accreting galaxy   must be updated  to become the starting model for the next merger. This is given by the transformations

\begin{eqnarray}
M_A        &=&  M_{C}        \nonumber  \\
M_{g,A}(j) &=&  M_{g,C}(j)   \nonumber  \\  
M_{s,A}(j) &=&  M_{s,C}(j)   \nonumber  \\ 
R_{s,A}    &=&  {R_{s,C}(j)} \\  
Z_A(j)     &=&  Z_C(j)       \nonumber  \\ 
<Z_A(j)>   &=& <Z_C(j)>      \nonumber  \\
l_{\Delta\lambda_{\gamma},A}(j) &=&  L_{\Delta\lambda_{\gamma},C}(j)/ M_C   \nonumber
\end{eqnarray}

\noindent
where   $j=1,....,281$, $k=1,....n_k$, and $\gamma=1,...n_{\Delta\lambda}$.          

vi) The radius $R_e$ and velocity dispersion $\sigma$ cannot  be evaluated in the same way. We follow here two different methods: (a) Given the mass of the galaxy, we make use of an empirical mass-radius relation. In literature, there are many possible relations mainly for elliptical and dwarf galaxies 
\citep[see for instance][and references]{Chiosi_etal_2020}. From this study we take the mass-radius relation of \citet{Fan_etal_2010}.
In the context of the $\Lambda$-CDM cosmology, \citet{Fan_etal_2010}  derived  an expression  (linking together (a) the halo mass $M_{DM}$, (b) the  stellar mass $M_s$ of the galaxy born inside it, (c) the half light (mass) radius $R_{e}$ of the stellar component, (d) the redshift at which the collapse took place $z_f$, (e) the shape of the baryonic component of a  galaxy (via the coefficient $S_S(n_S)$ related to the S\'ersic brightness profile from which the half-light radius is inferred and the S\'ersic index  $n_S$), and (f)  the velocity dispersion of the baryonic component with respect to that of dark matter expressed by the parameter $f_\sigma$, that is, $\sigma_s=f_\sigma \sigma_{DM}$, and, finally, (g) the ratio  $m=M_{DM}/M_s$.  The expression for the mass-radius relation  is:

\begin{equation}
R_{e}=0.9 \left(\frac{S_S(n_S)}{0.34} \right) \left(\frac{25}{m}\right) \left( \frac{1.5}{f_\sigma} \right)^2 
\left( \frac{M_{DM}}{10^{12}  M_\odot} \right)^{1/3} \frac{4}{(1+z_{f})},
\label{mr3}
\end{equation}

\noindent
Once the radius is known, we can derive the velocity dispersion  by means of 
\begin{equation}
 \sigma^2  =  (G M_s) /(k_v R_e),
 \label{sigma}
\end{equation}

\noindent 
where $G$ is the gravitational constant, and $k_v$ a term that gives the degree of structural and dynamical non-homology. The presence of $k_v$ allows us to write $M_s$ (stellar mass) instead of the total mass $M_T = M_{DM} + M_b$. $k_v$ is  a function of the S\'ersic index $n_S$, that is $k_v=(73.32/(10.465+(n_S-0.94)^2)+0.954)$ \citep[see][for all details]{Bertin_etal_1992, Donofrioetal2008}. For the sake of simplicity we assume $n_S=4$ as the mean value independently of the morphological type.
(b) Alternatively, radius and velocity dispersion can be derived from the ratios 
$(\sigma_f / \sigma_i)^2$ and  $R_{e,f}/R_{e,i}$  of eqns. \ref{Var_Ref_Gal_a} (where $i$ and $f$ indicate the stages before and after the merge, respectively) using a suitable definition of the parameters $\eta$, $\epsilon$ (based on the masses and velocity dispersion of the galaxies A and B), and $\lambda$ for which we assume that it can expressed as a (small) fraction of the velocity dispersion of galaxy A or B, and $\theta$ (which is from the luminosities of galaxy A and B).  In this case, the Virial Theorem is explicitly taken into account when sharing the total energy of the system to disposal between the kinetic  and gravitational potential  \citep[see][and references]{Donofrio_Chiosi_Brevi_2025}.

vi) Finally, some comments on the real path followed by a galaxy across the space of parameters $M_T$, $M_s$, $M_g$,  $Z$,  $<Z>$, $R_e$, $L_{\Delta\lambda}$, and others. Given a sequence of $N_k$ mergers supposed to occur to galaxy A by capturing at the ages $t_A$  a series of galaxies B of age $t_B$ (one per event), the path of the galaxy A across the parameter space of parameters   has to be extracted from the N sub-tabulations, ordered according the merger index $N$ and age index $j$, that we have calculated and stored. In other words, the path is the sum of all elemental partial paths  from the galaxy formation stage to the present via the various merger episodes. In general, the galaxy properties  smoothly vary according to their natural evolution, but at mergers they may suffer large changes on a very short time scale compared to galaxy lifetimes. In brief, the total path is the sum of the partial paths of different length connecting two successive mergers

$$ P(t) = \sum_0^{N_k} \Delta p_n(t) $$

\noindent
where $n$ varies from 0 to $N_k$ and $n=0$ stands for the initial part of the evolutionary history of galaxy A before the first merger. At each future mergers occurring at the ages  $t_{A,1}$, $t_{A,2}$, $t_{A,3}$ ..... $t_{A, N_A}$ (associated grid numbers $N_{A,j}$ in the age-grid $j$ (from 1 to $N_M$), the elemental path $\Delta p_n((t)$ is identified by the ages $t_{A,j}$ and $t_{A,j+1}$.

\section{Results from semi-analytical mergers }\label{sec:4_resana}
In this section we present some results from   the  semi-analytical  mergers. The aim is to test their ability as proxies of the data of real galaxies and/or the results of detailed numerical simulations. In particular, (i) we present the temporal evolution of five basic physical parameters, $L$, $M_s$, $R_e$, $\sigma$, and $I_e$;  (ii)  we   focus on the three  planes that can be set up with those physical quantities: namely  $I_e-R_e$, $L-\sigma$, and $R_e-M_s$; and finally present our  best description of  the $I_e$-$R_e$ plane based on the Virial Condition. 
To this purpose we calculated three groups of models according to the number of mergers that are supposed to occur during a galaxy's life, that is $N_k$= 5, 10,  and 20. 
Furthermore, for each value of $N_k$, we grouped models according to the maximum mass $M_{A,u}$ that can be reached by the galaxy A during the $N_k$ mergers. Equally for the mass $M_{B,u}$ of the merging galaxy $M_B$. In brief, for each case, the initial mass $M_{A,0}$ is randomly chosen in the range $5 \leq \log M_A \leq \log M_{A,u}$. Similarly for  $M_B$ the mass of which at each event is randomly chosen  between a constant lower limit of $10^5 M_\odot$ and $M_{B, max}$). Finally, no constraint is imposed on $M_B$ with respect to $M_A$.  The present formalism does not care whether $M_A \geq M_B$ or $M_B \geq M_A$.  
In general the accreting mass $M_B$ is lower than $M_{B, max}$. Consequently, for a number of mergers $N_k$, the limit mass  $M_{B, max}$ can be used to get an idea of the mass reached by the mass $M_A \simeq M_{A,0} + N_k M_{B,max}$ after $k$ mergers.  Therefore $M_{B,max}$ can be safely used as a parameter ranking galaxies according to the mass the can reached. The limit mass $M_{B, max}$ is shortly indicated by  $\tilde M$ and therefore used to group the galaxy according to $\tilde M$.
Throughout the paper we will consider groups of galaxies whose  $\tilde M$ goes from $10^6$ to $10^{13} M_\odot$ in steps 10. Finally, according to convenience we will either show results for all values of $\tilde M$ or limited to a smaller subset (typically   $\tilde M = 10^8$,  $10^{10}$, $10^{12}$, and $10^{13}$ $M_\odot$).

\begin{table}
\caption{Masses and ages of a typical merger sequence of  a galaxy with intermediate mass and dimensions. Masses are in $M_\odot$ and ages are in Gyrs.  }             
\label{table:mass_age_grid}      
\centering                           
\begin{tabular}{c c r c c r c}        
\hline\hline                 
 $n_k$ & $M_A$ & $t_A$ & $N_j$ & $M_B$ & $t_B$ & $N_j$ \\ 
 \hline
 \multicolumn{7}{c}{Number of mergers $N_k$=10}\\
\hline 
    0  & 3.59E+10  &  3.06 & 198  &           &        &      \\
    1  & 3.59E+10  &  3.06 & 198  &  1.68E+09 &  2.90  & 197  \\
    2  & 3.76E+10  &  4.08 & 205  &  2.02e+06 &  2.73  & 196  \\
    3  & 3.76E+10  &  4.35 & 207  &  5.82E+07 &  2.55  & 195  \\
    4  & 3.77E+10  &  4.35 & 207  &  6.78E+10 &  1.33  & 187  \\
    5  & 1.05E+11  &  4.48 & 208  &  3.80E+07 &  3.37  & 200  \\
    6  & 1.05E+11  &  6.24 & 222  &  1.45E+06 &  2.37  & 194  \\
    7  & 1.06E+11  &  7.19 & 230  &  5.90E+07 &  3.06  & 198  \\
    8  & 1.06E+11  &  8.91 & 245  &  9.30E+09 &  5.88  & 219  \\
    9  & 1.15E+11  &  9.14 & 247  &  3.72E+08 &  6.12  & 221  \\
   10  & 1.15E+11  & 12.00 & 273  &  3.86E+07 &  7.42  & 232  \\
\hline   
\end{tabular}                    
\end{table}

\subsection{Temporal evolution of  $M_s$, $R_e$, $L$, and $\sigma$ }

\begin{figure*}        
   \centering
   {\includegraphics[scale=0.22]{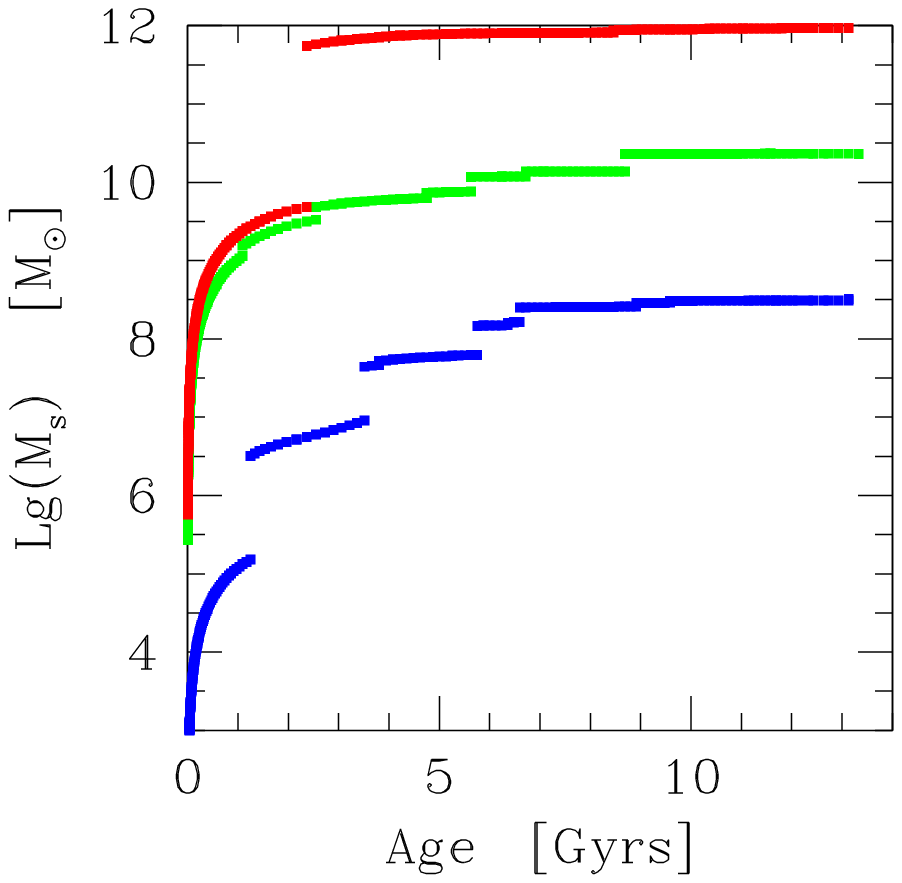}
    \includegraphics[scale=0.22]{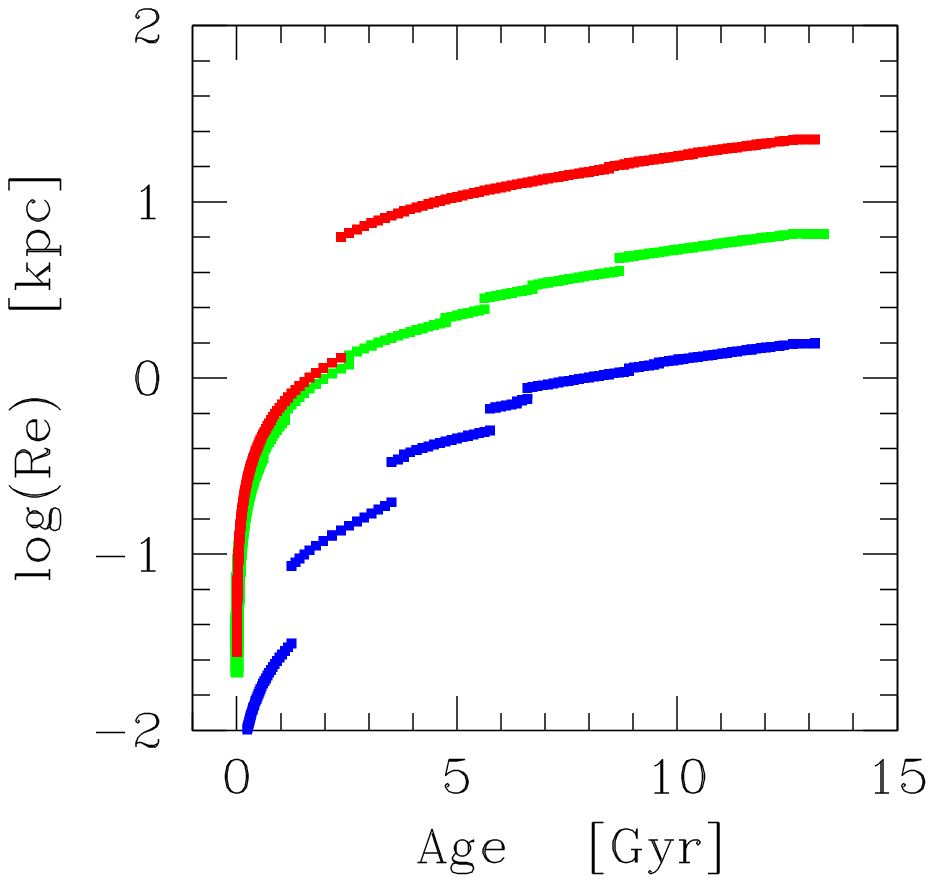}
    \includegraphics[scale=0.22]{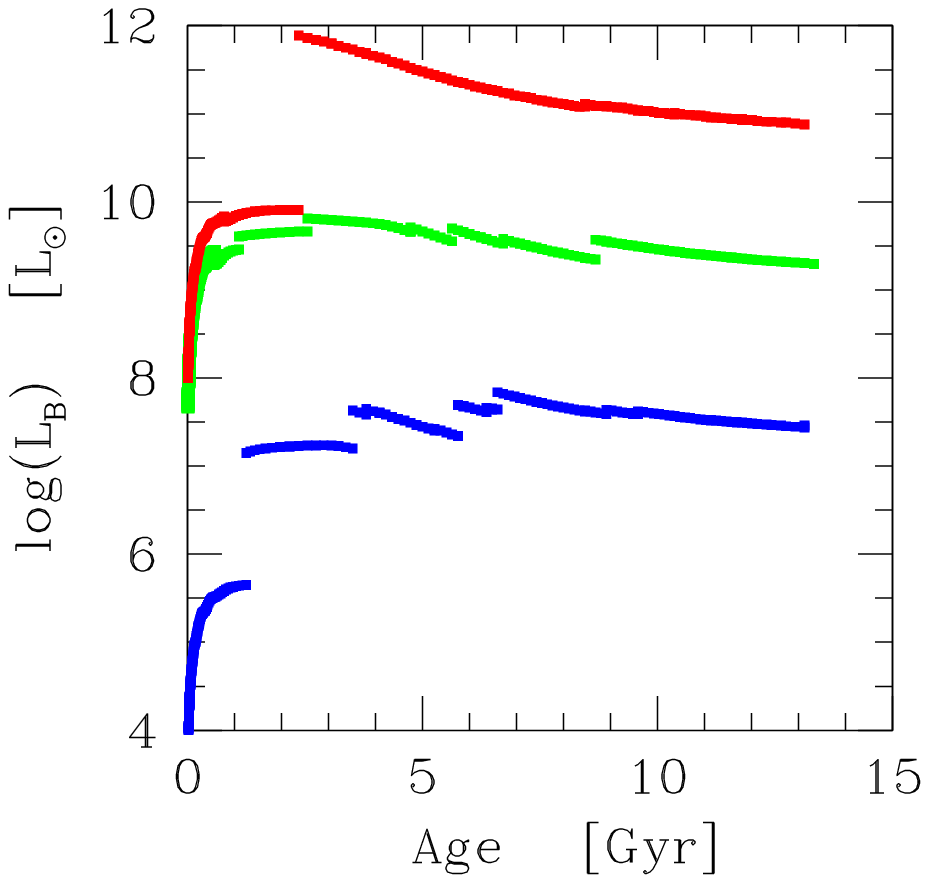}
    \includegraphics[scale=0.22]{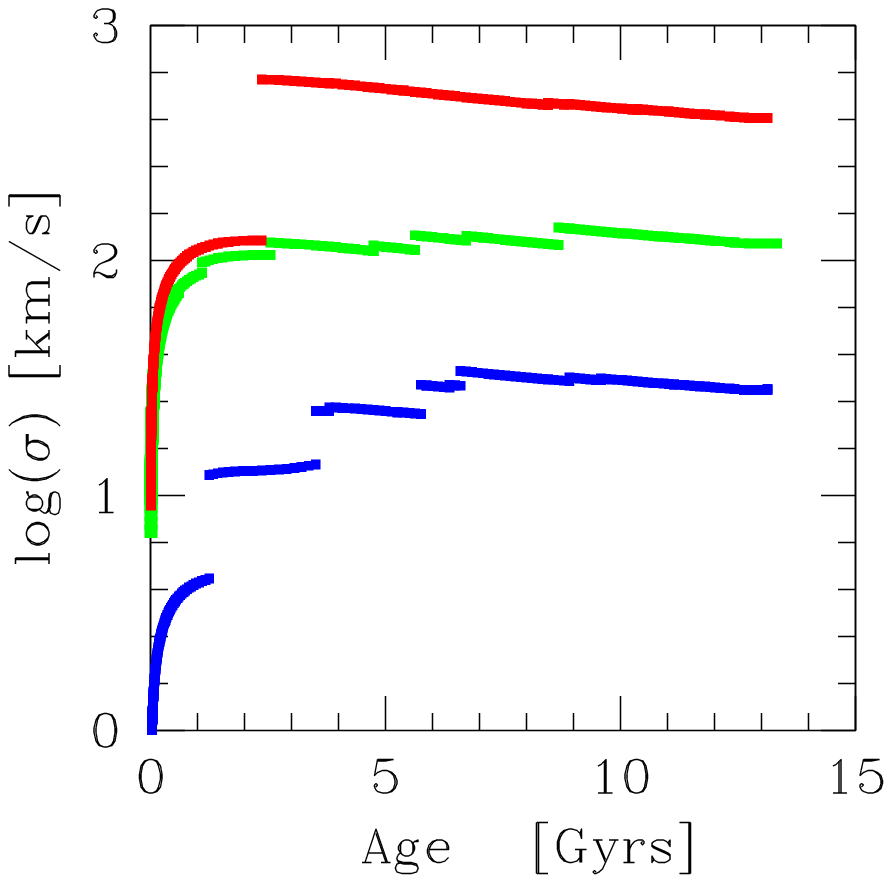}
    }
   \caption{ The age dependence in the galaxy rest-frame of four quantities of interest for three galaxies belonging to the 20 mergers group with $\tilde M =10^{8}$ (blue), $\tilde M =10^{10}$ (green)  and $10^{12}\, M_\odot$ (red). The age is in Gyrs. The quantities are:   mass $M_s$ in solar units ({ Left Panel}); effective radius $R_e$ in kpc ({ Middle Left Panel}); stellar luminosity  $L_B$ in the B-band in solar units ({ Middle Right Panel}); finally, the velocity dispersion $\sigma$   in km/s ({ Right Panel}). In all  panels, each curve is drawn in separate steps to better show the series of mergers that took place  along the whole path of a galaxy.
   }
              \label{Fig:Ms_Rs_Lb_sig_Age}
    \end{figure*}

In this section we present a preliminary qualitative analysis of the gross features characterizing these model galaxies. We focus  our attention on the time dependence of the physical parameters $M_s$, $R_e$, $L$, and $\sigma$ along the path of the model galaxy describing the sequence of mergers. To this aim,  
in Fig.\ref{Fig:Ms_Rs_Lb_sig_Age}  we show the temporal variation of the stellar mass 
$M_s$ (left panel, masses in $M_\odot$, ages in Gyrs),  the effective radius $R_e$ (middle left panel, radii in kpc, ages in Gyrs), the luminosity $L_B$ (middle left panel, luminosity in $L_\odot$, ages in Gyrs) and finally the velocity dispersion $\sigma$ (right panel, velocity in km/s, ages in Gyrs). The models under consideration belong to the group $N_k$=20 and $\tilde M=10^8, 10^{10}, 10^{12} \, M_\odot$, the ages at which mergers occurred can be estimated from the data displayed in Fig.\ref{Fig:Ms_Rs_Lb_sig_Age}.  Other choices for the age sequence of mergers would yield similar but not identical results.

As expected, the mass $M_s$ and radius $R_e$ increase as a functions of time in discontinuous steps according the amount of mass acquired at each merger. In the panels of Fig.\ref{Fig:Ms_Rs_Lb_sig_Age} for the mass $M_s$ (left panel) and radius $R_e$ (middle left panel), the solid lines correspond to the initial stages before the first merger.   
The same remarks and comments apply to the  middle right, and right panels of Fig.\ref{Fig:Ms_Rs_Lb_sig_Age} showing the variation of luminosity $L_B$ of the Johnson-Cousins photometric system (the luminosities for all other pass-bands are also available) and the variation of the velocity dispersion $\sigma$. 
In general, the temporal behaviour of the four relations can be easily understood in the frame of infall models of galaxies, the building blocks of our mergers among galaxies. In infall models of galaxies, $M_s$, $R_e$, $\sigma$ and $L$ rapidly increase with the timescale $\tau$,  and then level off as time passes.  The peak of star formation occurs at an age comparable to $\tau$ and then declines. This causes a steady decline of the luminosity, that is  exactly what we see in Figs. \ref{Fig:Ms_Rs_Lb_sig_Age}  together with the discontinuities caused by the mergers.

\begin{figure}        
  \centering
   \includegraphics[scale=0.45]{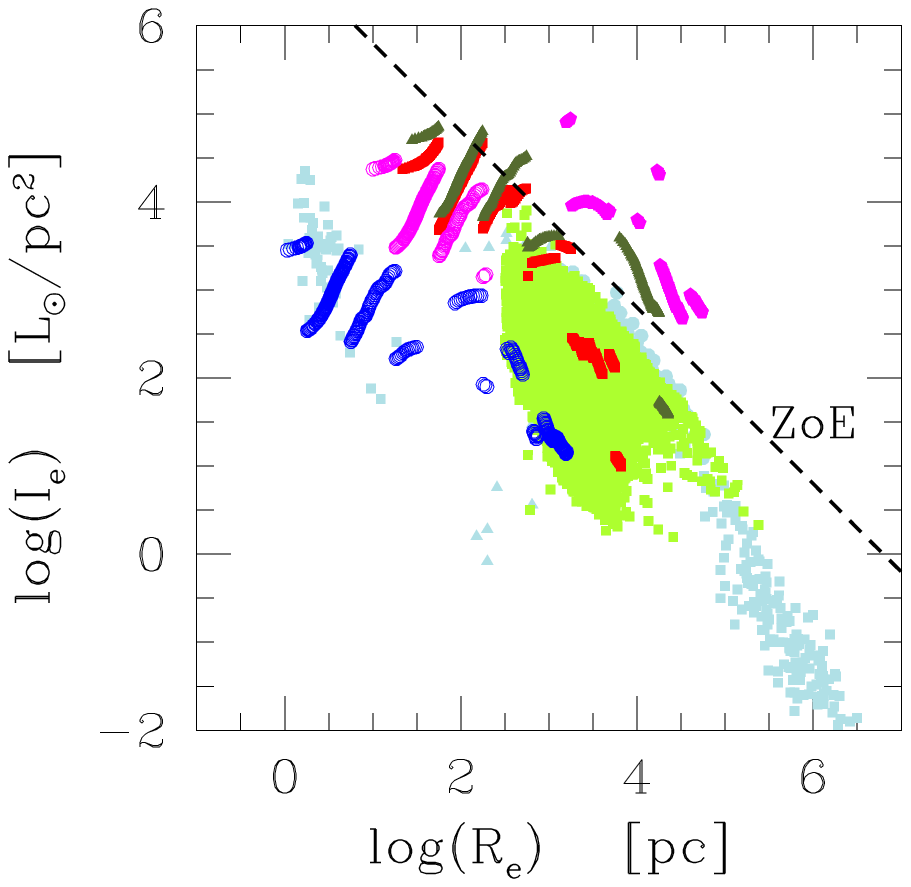}
   \caption{ The $I_e$-$R_e$ plane:   complete path of galaxy models with mergers belonging to the group $N_k = 20$. The galaxies on display have $\tilde M$ equal to $10^8$ (blue), $10^{10}$ (red), $10^{12}$ (dark-olive-green), and $10^{13}$ (magenta) $M_\odot$  from the redshift of galaxy formation $z_f=10$ to the present $z=0$. The various segments along each path from the start to the present visualize the complexity of the path followed by each galaxy. The comparison is made with the WINGS data (green-yellow points)  and the classical \citet{Burstein_etal_1997} sample (powder-blue points).} 
              \label{Fig:Ie_Re_plane}
\end{figure}

\subsection{Ie-Re plane}
  
\begin{figure}        
  \centering 
  {
  \includegraphics[scale=0.45]{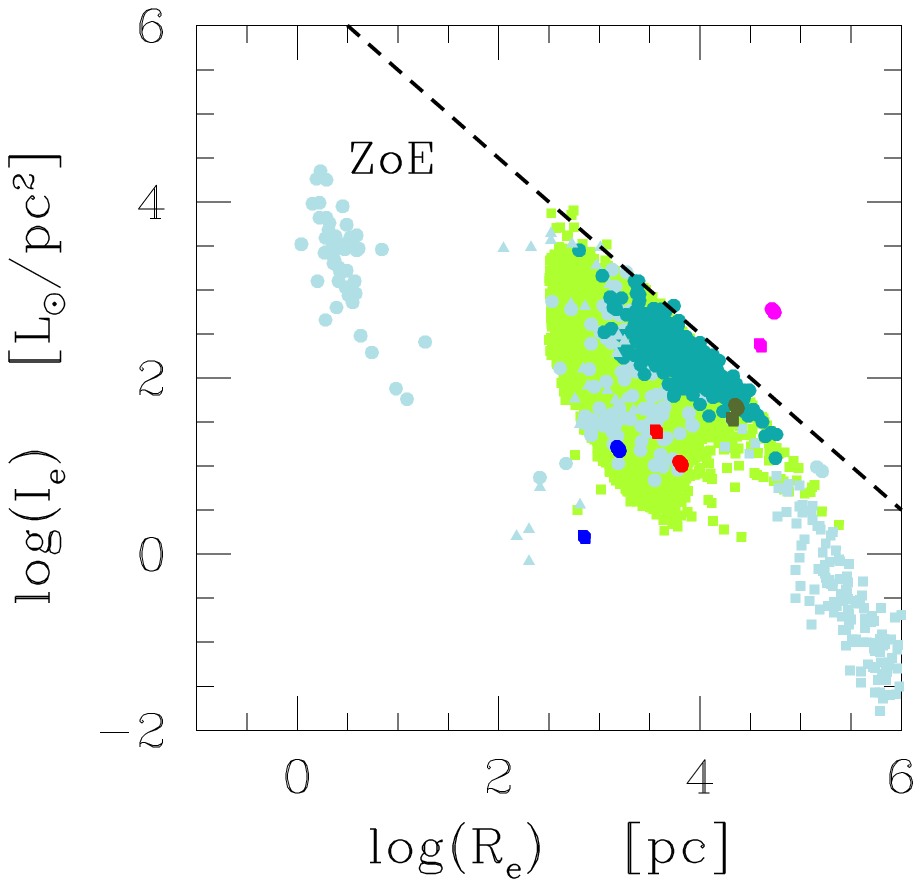}
   } 
   \caption{ The $I_e$-$R_e$ plane for galaxy models belonging to the groups with 20 (filled circles)  and 5 (filled squares) mergers. Only models at $z=0$ (present age) are plotted. The galaxies on display have $\tilde M$ equal to $10^8$ $M_\odot$ (blue), $10^{10}$ $M_\odot$ (red), $10^{12}$ $M_\odot$ (dark olive-green), and $10^{13}$ $M_\odot$ (magenta).   The  comparison is made with the WINGS data (green-yellow points)  and the classical sample of \citet{Burstein_etal_1997}  going from globular clusters to galaxy clusters (all data  in pale sky-blue but for the ETGs in dark gray). }
   \label{Fig:IeRe_Z0} 
\end{figure}

\begin{figure}        
  \centering 
  {
  \includegraphics[scale=0.45]{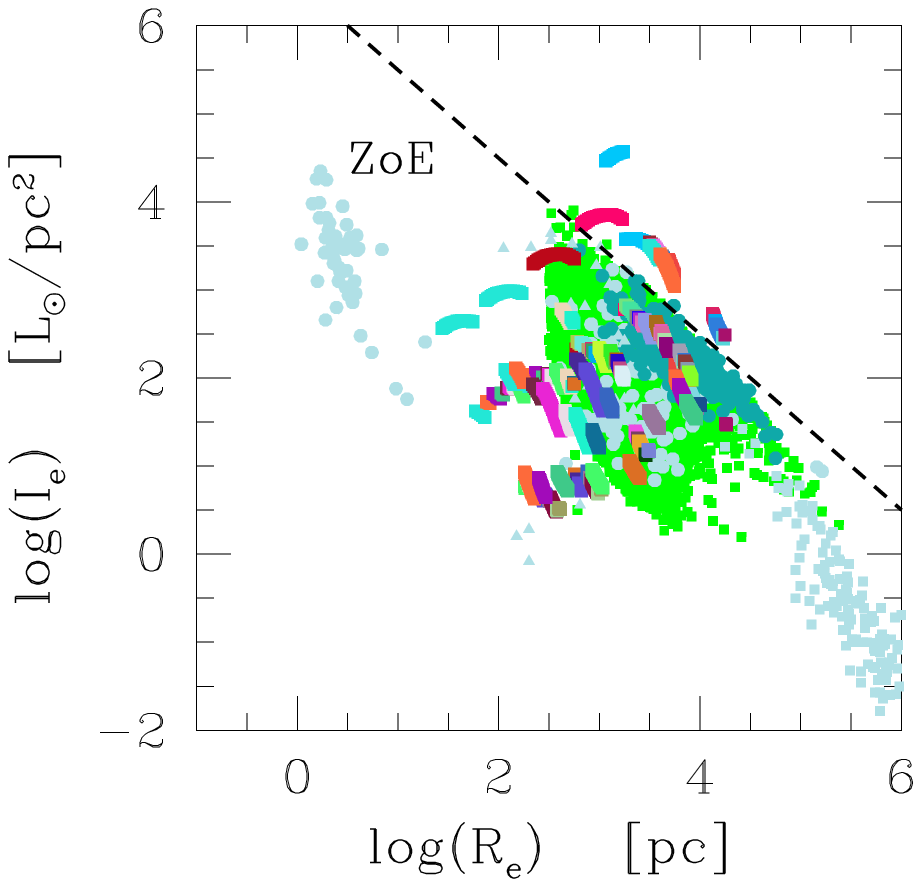}
   } 
   \caption{ The $I_e$-$R_e$ plane for galaxy models  belonging to the groups with 20   and 5 mergers (filled squares) mergers.  The galaxies on display have $\tilde M$ equal to $10^8$ $M_\odot$ (blue), $10^{10}$ $M_\odot$ (red), $10^{12}$ $M_\odot$ (coral), and $10^{13}$ $M_\odot$ (magenta).  We plot models from the first merger to the present time. Therefore these models span a wide range of possible ages.  The portions of each path from the start to the first merger and in between subsequent mergers are indicated by different colors. The  comparison is made with the WINGS data (green points)  and the classical sample of \citet{Burstein_etal_1997}  going from globular clusters to galaxy clusters (all data  in pale sky-blue but for the ETGs in dark gray). }
   \label{Fig:IeRe_agespan} 
\end{figure}

The $I_e$-$R_e$ plane is one of the key tests that theoretical models of galaxies have to pass. This is shown in Fig.\ref{Fig:Ie_Re_plane}, where we contrast the observational data of WINGs implemented by those of those of \citet{Burstein_etal_1997}  against our theoretical models of galaxies with mergers. The models on display are those with $N_k$=20 and are tagged by the color code as follows: $\tilde M = 10^8$ (blue), $10^{10}$ (red), $10^{12}$ (dark olive green), and $10^{13}$ $M_\odot$ (magenta). The lists of the merger ages are not of interest here. For each case, we show the whole path.  Because the observational data refers to  galaxies in the local Universe at the present time, it goes without saying that the comparison with the observational data should be limited only to the final stage of each path. However, the main purpose of Fig.\ref{Fig:Ie_Re_plane} is  to give  an idea of the complicated path followed by  a galaxy undergoing repeated mergers during its lifetime. The general trend shared by each sequence is first to proceed at increasing $I_e$ and $R_e$  and then at decreasing $I_e$ and increasing $R_e$. This roughly mimics the variation of the luminosity with the star formation  activity: at early epochs the star formation activity increases with time, the mass of the galaxy increases by infall and mergers, finally the luminosity and $R_e$ increase. The radius increases mainly because the mass increases. When star formation ceases, the luminosity tends to declines, although  successive mergers tend to keep it at high values, later  the  increase of $R_e$  by mergers prevails and $I_e$ goes down.  The path of a galaxy during this later phase tend to follow the slope of the ZoE. 
The comparison between data and theoretical models, however limited to the present time or very recent past (redshift $z < 0.03$), is  shown   in  Fig. \ref{Fig:IeRe_Z0} using the same models with $N_k=20$ (filled circles) shown in Fig.\ref{Fig:Ie_Re_plane} together with also models of the group $N_k=5$ with the same mass $\tilde M$ and redshift interval (filled squares). As expected, the models still fall into the region of the observational data but may account only for a limited number of objects, that is the oldest galaxies crowding at the low luminosity (low $I_e$) border of the distribution. What about the remaining objects? We propose the following explanation. The theoretical models with mergers we have considered are based on the hypothesis that all galaxies  are  born at $z_{for} = 10$. This is a severe limitation that should be relaxed, in particular for the low mass objects. It follows from it  that in our observational sample there could be galaxies significantly younger than the canonical age of 13.187 Gyrs ($T_{U,0} - T_{U,f}$ = 13.671 - 0.484 = 13.187 Gyrs ). To a first approximation, without repeating all the calculations, the point can be partially cured by plotting also objects at younger ages. This is what is shown in Fig.\ref{Fig:IeRe_agespan} where model galaxies with ages going from a few Gyrs to the present are plotted all together. The coverage  of observational data is better than before. However, the result cannot be taken as final because correct  models with $z_f$ spanning a wider range should used in particular for the low mass galaxies. Work is in progress to include this option.

Figure \ref{Fig:IeRe_agespan} indicates that the hierarchical scenario in which galaxies  are built up by successive mergers is compatible with the observational data. The main body of data can be matched by galaxies of different age that underwent a number of mergers during their lifetime. However, the particular shape of the low-$I_e$ limit and the tail of objects (ETGs) running along the ZoE boundary and merging the longer tail of galaxy groups and clusters require some comments. The $I_e$-boundary is the locus of old objects of increasing mass, luminosity, and radius that  can be explained even by simple infall models of galaxies, no matter whether mergers have occurred or not \citep[see the model by][]{Donofrio_Chiosi_2023a}; mergers play a minor role. 

Another remarkable feature to note are the early stages of the models with $\tilde M$ in excess of $10^{10} M_\odot$ that fall into forbidden regions of the plane according the present day ZoE. They are young objects like those expected at high redshift (say above $z = 1$). Their existence should be proved by high redshift data \cite[see][for a preliminary analysis of this topic]{Donofrio_etal_2024}.  Furthermore, there are models with ages old enough to be virtually present in the local universe that also fall beyond the ZoE (some of the magenta squares), their mass is above $10^{13} M_\odot$. This could imply that  galaxies with this mass cannot form or are extremely rare.

The tail of ETGs running along the ZoE toward larger and larger radii and fainter $I_e$'s is not easily  reproduced by our models. The problem is not with the number of mergers but with the type of merger that should occur. That is  a merger between a massive, quiescent galaxy on which of objects of small mass are captured by merger. The large mass difference between the two galaxies should secure that the luminosity of the host object does not change in a significant way.  This statement  stands on simple common sense considerations: in a given volume of space an objects cannibalizing its neighborhoods finds more and more difficult to merge one of comparable size. This issue should be considered when randomly choosing the mass $M_B$ of the galaxy to merge with an existing object of mass $M_A$ and old age. Work is in progress to include this effect in the numerical code. 
At present we limit ourselves to say that old massive objects with low $L$ (no active star formation) and large  $R_e$ are the best candidates to fall in this particular region of the $I_e$-$R_e$ plane,  see also \citet{Donofrio_Chiosi_Brevi_2025}  for models of this type but based on the simple analytical description of mergers with the aid of the Virial Theorem.

Furthermore, one could also argue that  the average number of possible mergers in galaxies of different mass  could vary with the mass:  a large number of mergers across the whole galaxy age for the more massive objects and fewer  mergers for the low mass ones.  The choice of the number of mergers is based on simple theoretical arguments of statistical nature.

\subsection{On using the virial condition}

In the model galaxies with mergers we have presented in the previous section, the radius $R_e$ and the velocity dispersion $\sigma$ were calculated with our first method:  
once a merger has occurred, the star mass $M_s$ is the sum  $M_{s,A} + M_{s,B}$  and the new value is used to derive from relation (\ref{mr3} of \citet{Fan_etal_2010} the new radius $R_e$. With the new $M_s$ and $R_e$,  we derive from eqn.(\ref{sigma}) the new value of $\sigma$. In this case, the Virial Theorem is only indirectly taken into account via eqn.(\ref{sigma}).  

\begin{figure}        
   \centering
   \includegraphics[scale=0.45]{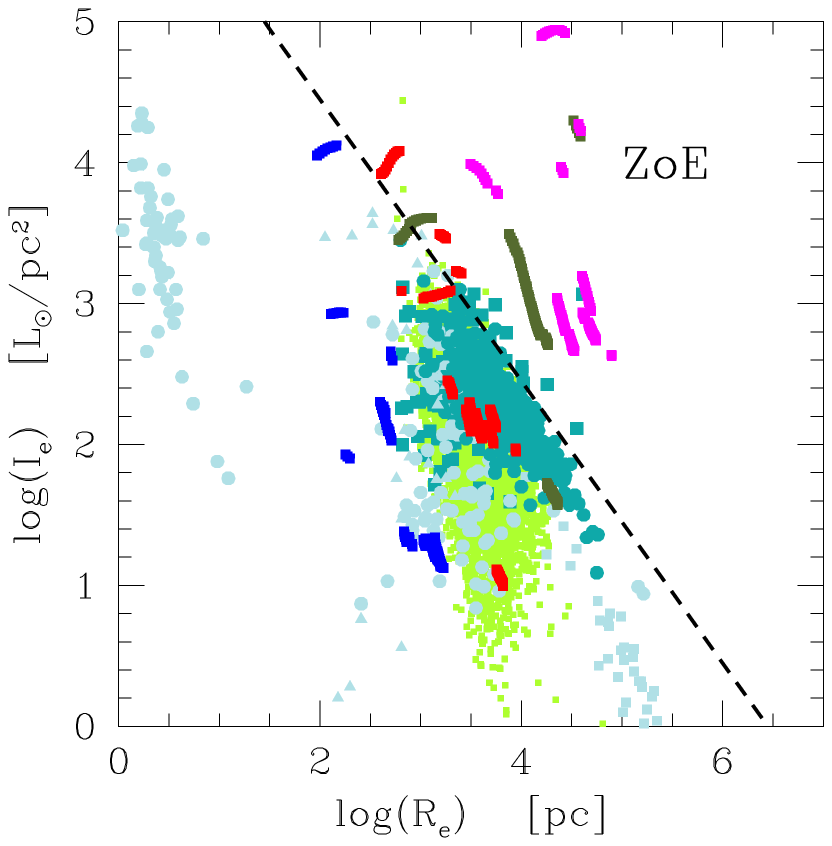}
   \caption{The $I_e$-$R_e$ plane of models whose radius and  velocity dispersion are derived imposing the Virial Condition according to the formalism of eqns. (\ref{Var_Ref_Gal_a}).  The difference with respect to the previous cases is small, although not negligible). We consider them as the ones to use.}
   \label{Fig:Ie_Re_virial}
    \end{figure}

Alternatively our second method can be used, that is radius and velocity dispersion  are derived from the relation 
$(\sigma_f / \sigma_i)^2$ and  $R_{e,f}/R_{e,i}$  of eqns. (\ref{Var_Ref_Gal_a}) (where $i$ and $f$ indicate the stages before and after the merge, respectively). This requires  a suitable definition of the parameters $\eta$, $\epsilon$, $\theta$, and $  \lambda$.  They are given by $\eta = M_B /M_A$  and 
$\epsilon = \sigma_B / \sigma_A$, $\theta= L_A/L_B$,  where $M_A$, $M_B$,  $L_A$ and $L_B$ are updated at each merger.  As far as  $\lambda$ is concerned, we assume   $\lambda = 0.0001 (V_{kin} /\sigma_A)^2$ where $V_{kin}$ is the relative velocity between the two galaxies, $\sigma_A$ is the velocity dispersion of the galaxy A, and $V_{kin} \simeq 500$ km/. The factor 0.0001 is meant to keep the effect very small.  In this  case, the Virial Theorem is explicitly taken into account in the re-partition of the total energy between the kinetic and gravitational potential  \citep[see][and references]{Donofrio_Chiosi_Brevi_2025}.  The numerical calculations show that but for a period of time  soon after the merger, the new $R_e$ and $\sigma_A$ obtained with the two methods are nearly identical and consequently so is the path of model galaxies on the $I_e$-$R_e$ plane.  The new models are compared to observational data on the $I_e$-$R_e$ plane of Fig. \ref{Fig:Ie_Re_virial}. We consider the new plane $I_e$-$R_e$ just obtained as the best representation of the reality  because  radii and velocity dispersion are both directly derived from the virial relations.

\subsection{Fluctuations in the L-$\sigma$ plane}
\begin{figure*}        
  \centering
  {
   \includegraphics[scale=0.45]{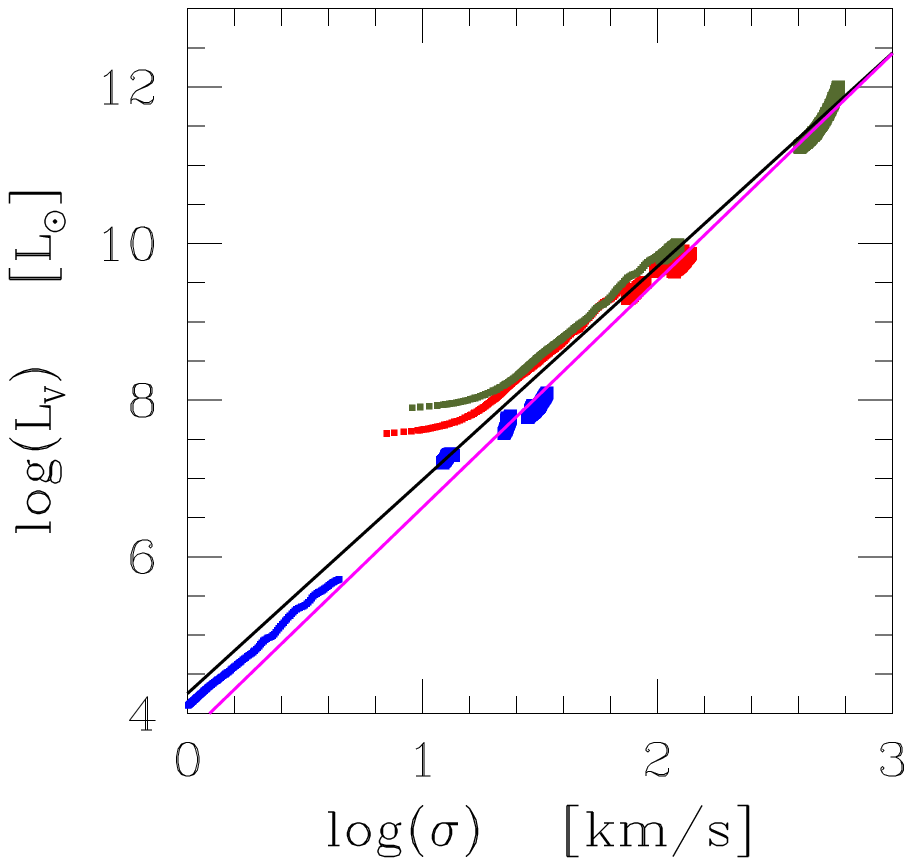}
   \includegraphics[scale=0.45]{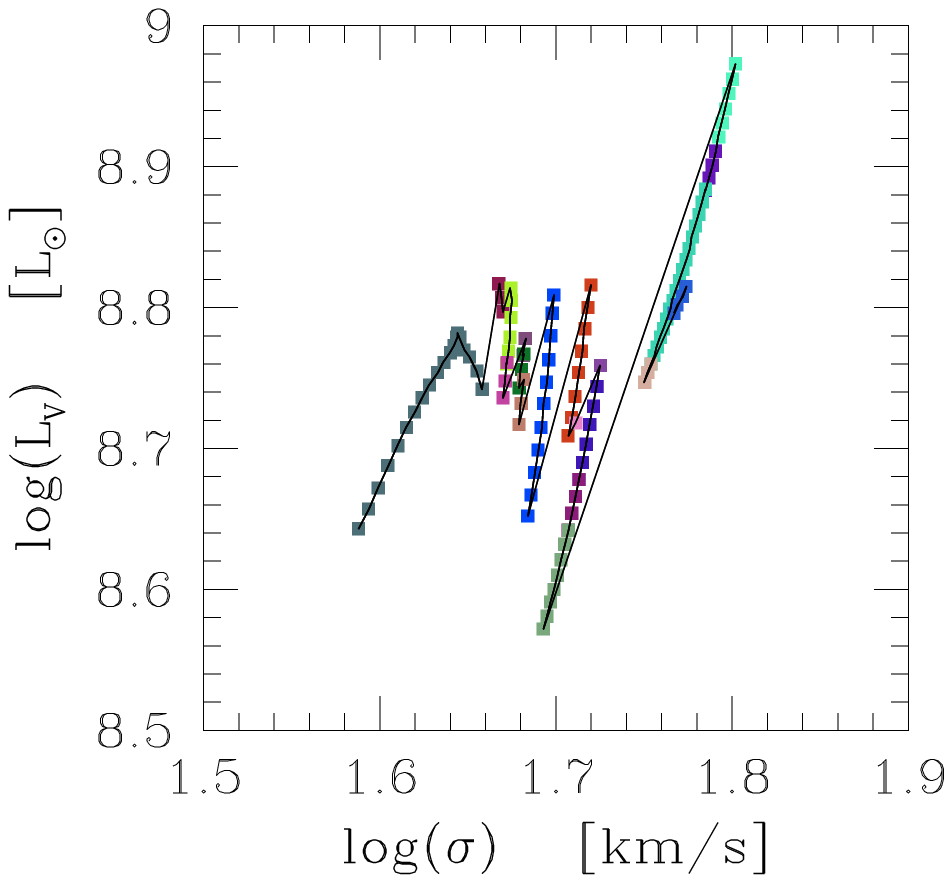}
  } 
   \caption{ {\bf Left Panel}: The $L-\sigma$ plane  of three model galaxies, namely  $\tilde M = 10^8\,  10^{10}\,  {\rm and}  10^{12}\, M_\odot$, undergoing 20 mergers. The complete paths are shown. The best fit of the models is shown by the solid lines (black if all evolutionary stages are included, magenta if the pre-merger stages are excluded).     
   {\bf  Right Panel}: blow-up of the path of the $\tilde M = 10\, M_\odot$ galaxy undergoing 20 mergers. Each portion of the path is made by squares of different color in order to better visualize the variation of direction in $L-\sigma$ plane. }
\label{Fig:Lum_sig}
\end{figure*}

In the following we want to go over the question of the real relation between luminosity and velocity dispersion, that is if there are  argument to support the idea that  the exponent $\beta$ can vary with time in a galaxy and/or from galaxy to galaxy. In the left panel of Fig.\ref{Fig:Lum_sig} we plot the luminosity $L_V$ (in the classical Johnson-Cousins pass-band) versus the velocity dispersion $\sigma$. The luminosity is in solar units and the velocity dispersion in km/s.  The data on display corresponds to the sequences of 20 mergers for galaxies whose masses $\tilde M$  are $10^8$ (blue), $10^{10}$ (red), and $10^{12}$ $M_\odot$ (magenta). In each sequence, the solid part corresponds to  the pre-merger stages, whereas the squares of different colors (as appropriate) indicate the portion of the path in between subsequent mergers. 
The best fit of all the data yields the following luminosity-velocity dispersion relationship  $\log L_V = 2.727 \log \sigma  +  4.253$ whose slope is in satisfactory agreement with current  observational data, 
the slope of which is about 3.5. The slope given by the theoretical models increases  if we exclude from the evaluation all models in the pre-merger phase which  refers to very initial stages that of course are not accessible to current  observations in galaxies of the local Universe. In this case, we get $\log L_V = 2.899  \log \sigma + 3.731$. The agreement between theory and observation has slightly improved.
In any case the overall $L-\sigma$ looks well behaved and linearly smooth as we move toward higher and higher masses, thus supporting the common sense notion of  \Lsigb\ relation with $\beta$ constant all over the mass range. However, if we have a closer look at the  $L-\sigma$ relation we note that because of the repeated mergers the relation is actually made by a numberless series of broken lines each one with its own slope and zero point, as shown in right panel of Fig. \ref{Fig:Lum_sig} for the particular case of the $\tilde M = 10\, M_\odot$ galaxy undergoing 20 mergers  (all other cases we have calculated show similar behavior). The mean \Lsigb\ is indeed made by a large number of individual \Lsigb\ that are best represented by the relation \Lsigbtempo\ with $\beta(t)$ and $L'_0(t)$ both time dependent quantities continuously  varying from galaxy to galaxy and also for the same object from stage to stage.

\begin{figure}    
   \centering
   {
   \includegraphics[scale=0.45]{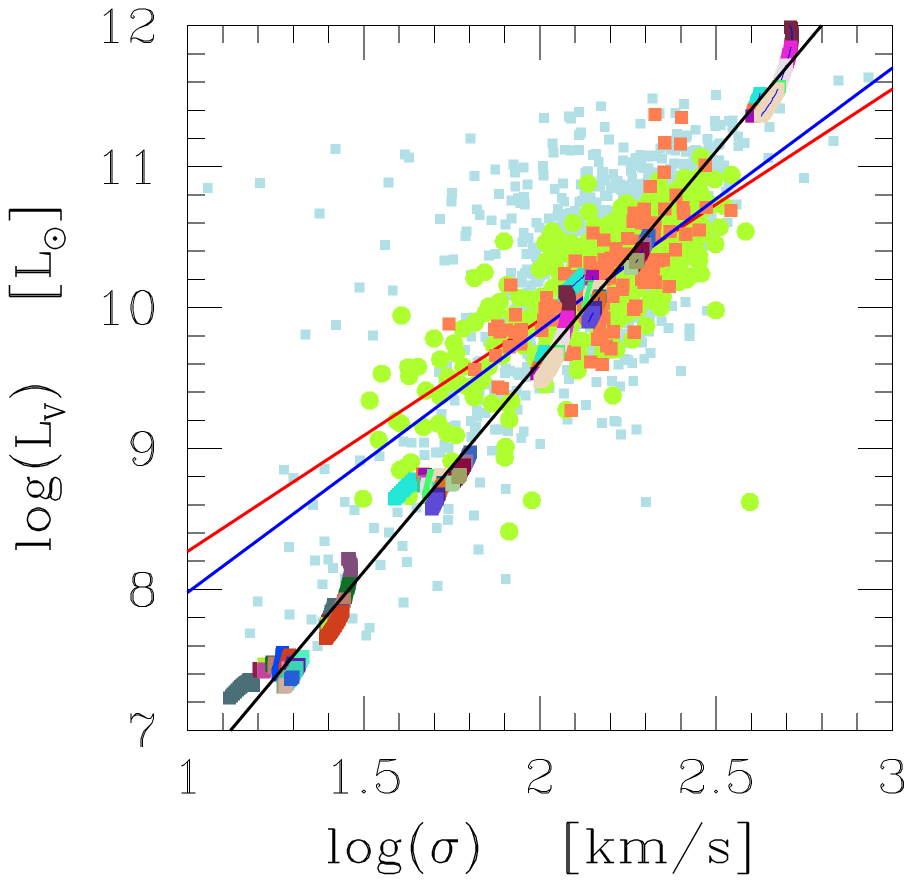} 
   }
   \caption{ Plane $L$ - $\sigma$ of galaxies with $N_k$ equal 20 and 10 (filled squares of different colors, no distinctions between the two groups)) and the observational data of MANGA (green filled circles), WINGS (coral filled squares) sand \citet{Burstein_etal_1997} (pale sky-blue dots).}
   \label{Fig:Lum_sig_2}
    \end{figure}

The second important relation to look at is  the  \Lsig\  shown in   Fig. \ref{Fig:Lum_sig_2} for the groups of models with $N_k$=10, and 20 mergers. The best fit of these data is 
$ \log L_V = 3.03 \log \sigma + 3.53$ ($L_V$  in $L_\odot$ and $\sigma$ km/s) as indicated by the solid black line.
To compare theoretical results  with observations we added the MANGA data (green filled  circles), the WINGS data (coral filled  squares), and in the background the \citet{Burstein_etal_1997} data (pale sky-blue dots) going from dwarf galaxies to galaxy clusters. Looking at data for ETGs only in MANGA (green filled circles) we get   $ \log L_V = 1.64 \log \sigma + 6.63$ and $ \log L_V = 1.86 \log \sigma + 6.11$ for WINGS. The slope is about a factor of two smaller. No best-fit of the  \citet{Burstein_etal_1997} data is made. The slope of the model best-fit is in perfect agreement  with the expectation from very simple considerations. 
In brief, if the velocity dispersion can be related to the  Virial Theorem and the radius $R_e$ in it   can be replaced by $R_e \propto M_s^{0.33}$ \citep[see][]{Tantaloetal1998},   to a first approximation, it follows that $\sigma^2 \propto M_s^{0.67}$. According to a typical mass to luminosity ratio for old galaxies, $L \propto M_s$. 
Therefore $L \propto \sigma^{3}$. The luminosity should only increase with $\sigma$. This is indeed the slope of the classical \Lsig\ relationships such as the Faber-Jackson.
However, we have seen that the \Lsig\ relation is more complicated than a simple linear dependence ($logL \propto log \sigma$) as shown in the right panel of Fig.\ref{Fig:Lum_sig}.
Schematically  two regimes are possible: at increasing mass the luminosity increases. However, for  a system in mechanical equilibrium, that is subjected to the VT, the velocity dispersion may either decrease, if there is not enough energy to disposal, or increase when the amount of the brought-in energy is large enough. In our models, given the mass of the captured object ($\eta$), there are two possible sources of energy expressed by the parameters, $\epsilon$ (the amount of internal energy brought-in by the captured object ) and $\lambda$ (the amount of kinetic energy brought-in by the captured object).  In general, the radius of the composite system is larger or equal to that of the hosting object. 
If there is little energy to disposal, in order to reach the equilibrium configuration, part of the      
energy must be taken from the dispersion velocity which may even decrease or remain constant. If $\epsilon$ and $\lambda$ are large enough, the energy required to reach the equilibrium state is taken from the kinetic energy of the in-falling object and the velocity dispersion may increase. It goes without saying that all intermediate situations are possible. This effect could explain the different slope (smaller by nearly a factor of two and the large dispersion seen in the observational \Lsiga\ relationship.

\subsection{$R_e$-$M_s$ plane}

\begin{figure}    
   \centering
   {
   \includegraphics[scale=0.45]{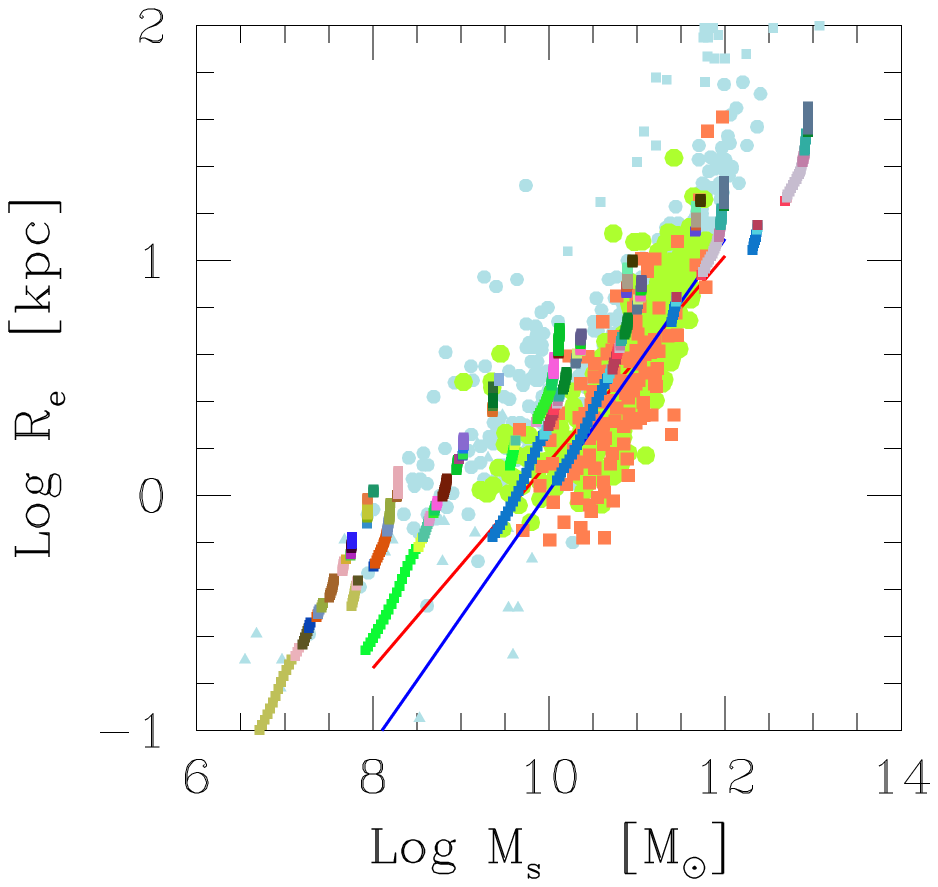} 
   }
   \caption{ $R_e$ - $M_s$ relation of galaxies with $N_k$ equal 20 and 10 (filled squares of different colors, no distinctions between the two groups)  and the observational data of MANGA (green dots), WINGS (red dots), and \citet{Burstein_etal_1997} (pale sky-blue dots).}
\label{Fig:MR_last}
    \end{figure}

There is not much to say about the $M_s$-$R_e$ relation shown in Fig.\ref{Fig:MR_last} together with the observational data, that is the MANGA (green dots) and WINGS data for ETG's (red dots),   also the data of \citet{Burstein_etal_1997} (pale-sky-blue dots), and 
the models of the 10 mergers group (colored filled squares).
The observational data of the MANGA data have best fit  $\log R_e = 0.44 \log M_s -4.26$ (solid blue line) and those from WINGS (red dots) have best fit $\log R_e = 0.54 \log M_s -5.35$ (solid red line).
Along each theoretical $M_s$-$R_e$ sequence we show the pre-merger stages (solid line) with a different color. In the theoretical relation only the portion after the first merger is plotted. All theoretical models have a good portion of their $M_s-R_e$ relations (the part corresponding to recent ages) falling inside the regions of the observational data. The large dispersion of the data can be accounted for.

The aim is to show that the theoretical $M_s$-$R_e$ relation of \citet{Fan_etal_2010} that we have adopted not only varies with the redshift but also  it can adjust itself to match the radius of objects at the extremes of the mass range going from globular clusters  (these latter only marginally indeed) to present day ETGs.  As already said, globular clusters are only marginally matched by the theoretical $M_s$-$R_e$ relation. However, in the context of our analysis this is less of a problem, because these systems have been dynamically reshaped by the interaction with the parent Milky Way. 
What we learn from the $M_s$-$R_e$ relations is that our models with mergers can match  the observational data in this plane.

\subsection{The star formation rate}
\begin{figure}        
   \center
   \includegraphics[scale=0.45]{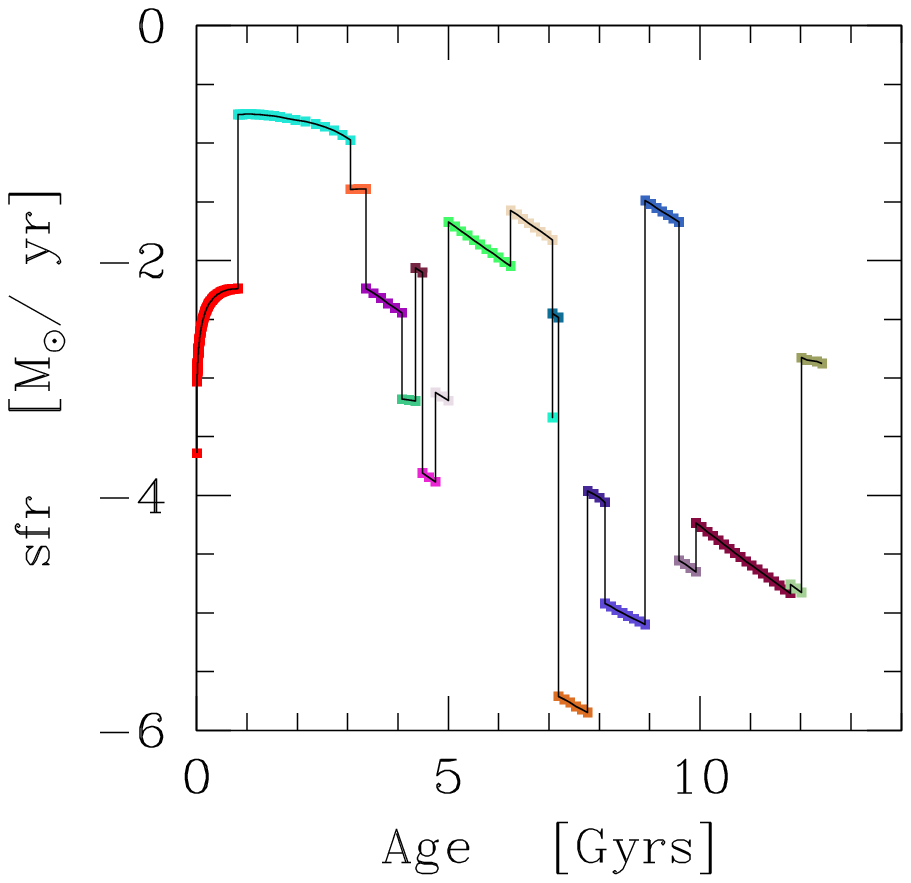}
   \caption{The SFR vs age in $M_\odot$/yr for a typical $\tilde M = 10^{10}\, M_\odot$ belonging to $N_k=20$ group. Note the appearance of pseudo-bursts of star formation in a model galaxy undergoing mergers. In absence of these latter a galaxy of the same mass would exhibit a ever smooth SFR starting small, increasing to a maximum at the age of about 1 Gyr, and then steadily declining to nearly zero at the present time. }
  \label{Fig:SFR_age_bursts} 
    \end{figure}

 An interesting feature of  the galaxy models with mergers is the  peculiar behavior of the star formation rate as a function of time. The typical time dependence of the SFR in a ``infall model'' of a galaxy is as follows: the SFR starts small grows to a maximum and then smoothly declines to nearly zero at the present  time. The age at which the peak occurs is about equal to the infall time scale $\tau$. The specific efficiency $\nu$, if assumed to be constant in time and independent of the galaxy time,  does not alter this trend. In this paper, the infall models are the seeds of the merger mechanism. Galaxies A and B before merging are described by an infall model of a certain mass. While galaxies B always obey the above scheme, galaxy A departs from it as soon as the first merger occurs. The explanation resides in the equation governing the SFR when a merger occurs: after merging the new SFR is the mass weighed mean of  $\rm SFR_A$ and $\rm SFR_B$ trend and consequently can depart from the usual trend.  As the number of mergers increases, the net results for the $\rm SFR_A$ is the oscillating curve shown in Fig.\ref{Fig:SFR_age_bursts}. We call it pseudo-bursting mode of star formation because in our mergers there is no additional star formation to distinguish it from the more realistic case in which at each merger a burst of new star formation most likely occurs (at least if there is gas left over and the merger is above some threshold level of efficiency).

\subsection{Final remarks on the models and their comparison with  observations}\label{sec:6_comp}  
  
In concluding this section, we add a few comments:   
 i) First of all, we emphasize that even if the cases presented here look alike as far as the general trend of the various relationship, in reality there are important differences that depend on the number of mergers that are supposed to occur and also on the properties of the recipient and captured galaxy at the time of merger, e.g. the age, the mass, and the time scale used for infall models. Since there are no specific arguments for preferring a particular model with respect to others, we expect  a large variety of  results to be  possible. This is indeed what happens in nature where mergers occur among  galaxies of different mass, radius, ages, chemical properties, luminosities, energy budget, evolutionary state,  and so forth.  
ii) The variations of luminosity caused by a merger are more difficult to predict. In a galaxy the luminosity is the sum of the partial luminosities emitted by stars of different generations and different chemical composition. Each generation contributes in a way proportional to its mass and age. The mass of a stellar generation is proportional to the rate of star formation at the time of its birth. In a single stellar generation, the luminosity decreases with the age because of its natural fading as the turnoff mass gets smaller and smaller. According to the analysis of the luminosity-age relationship of single stellar populations presented by \citet{Donofrio_Chiosi_Brevi_2025},  we note a fast decline during the first 1.5 to 2 Gyr after the start of star formation, followed by a smooth decline during the rest of the life by a factor of 3 to 5. In our case, we are dealing with the integrated luminosity-age relations of two model galaxies in which  the effect of different mean chemical composition are already taken into account,  we combine the two luminosity age relationships to obtain the luminosity-age dependence of the composite object, and we repeat the procedure merger after merger.  In absence of additional star formation at the merger of galaxies A and B, the resulting luminosity will be $L = (M_{s,A} L_A + M_{s,B} L_B) /(M_{s,A} + M_{s,B})$. In analogy, if a new generation of stars is created at the merger with total mass $M_{s,C}$ and luminosity $L_C$, the resulting new luminosity will be  $L = (M_{s,A} L_A + M_{s,B} L_B + M_{s,C} L_C)  /(M_{s,A} + M_{s,B} + M_{s,C})$. In our models, the  bursting star formation at the merger is missing. This is a feature of the model to be improved. 
iii) In the $I_e$ vs $R_e$ plane all the groups of models draw lines that are nearly parallel to the ZoE, confirming that they correspond to loci along which virial equilibrium is obtained (imposed). The scatter of the models along each line is due to the scatter of the properties of merging galaxies.  
iv) The $L$ versus $\sigma$ plane is intriguing. As expected the gross behavior is that the luminosity increases with the mass. However, looking at the particular history of a galaxy through its merger sequence, the possibility  arises that the luminosity increases while the velocity dispersion remains nearly constant or even  decreases. This is the result of the energy exchange among successive  mergers. Similar result is found and discussed by \cite{Naab_etal_2009} and \citet{Donofrio_Chiosi_Brevi_2025}.  In any case a large dispersion in this relationship is expected.
v) In the  $R_e$ versus $M_s$ relation, the radius always increases with the mass passing from a low mass to a high mass galaxy. In each individual galaxy,  the radius is expected to increase also  along the merger sequence. 
vi) To conclude, the theoretical models reasonably match the observational data  and can be considered as a proxy of the merger mechanisms of galaxy formation as far as masses and sizes and other quantities associated to these are concerned. 
vii)  Although the present theoretical models reasonably match the observational data, 
the agreement could improve if we knew the numerical frequency of objects of different masses,  the frequency of mergers between galaxies, and an estimate  of the number of mergers that each object of a certain mass can undergo.  It is reasonable to assume that in the low mass galaxies  only a limited number of mergers can occur, while in the most massive ones  a larger number of mergers is plausible. Finally ,the tail observed in $I_e$-$R_e$ towards larger and larger radii and lower and lower $I_e$'s  require merger of low mass objects of any age with massive old galaxies in which star formation extinguished long time ago.

\section{Our models in the context of the $\beta-L_0'$ theory}
\label{sec:5_betatheory}

In this section we present and discuss a profound correlation we have found between the $\beta$ parameter, the presence of mergers, and the condition of virial equilibrium. We consider this correlation as the most intriguing result of the  present analysis.

  \begin{figure*}        
   \centering {
   \includegraphics[scale=0.40]{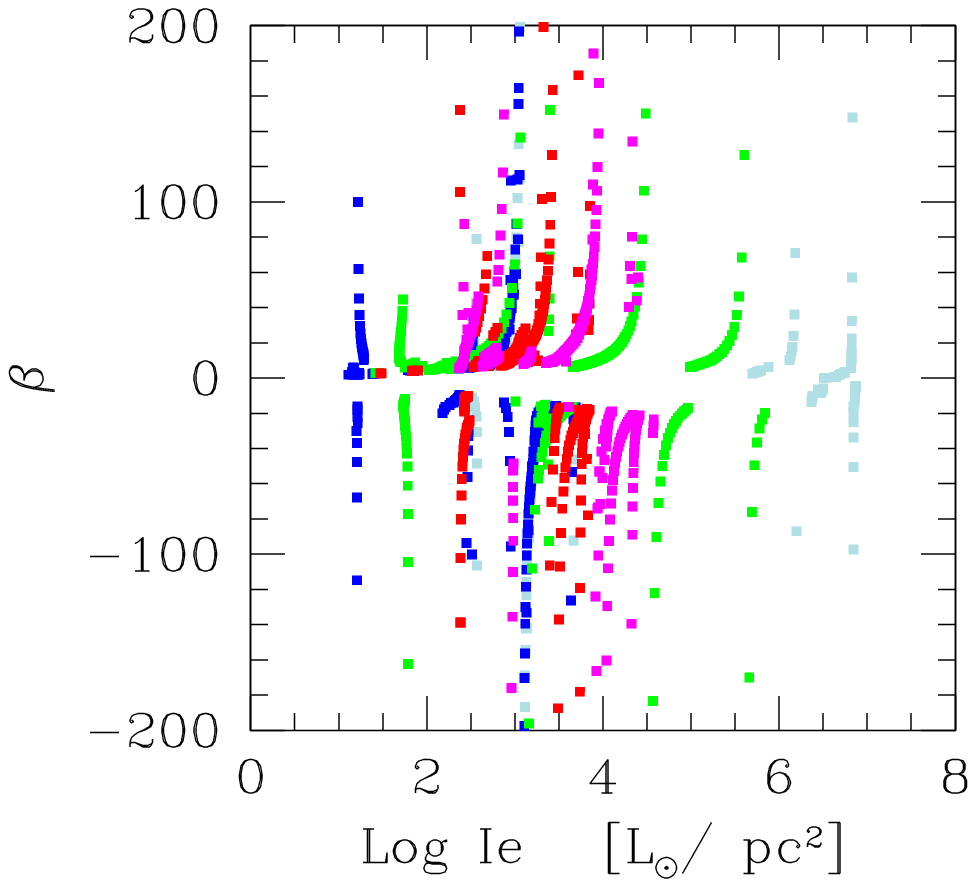} 
   \includegraphics[scale=0.40]{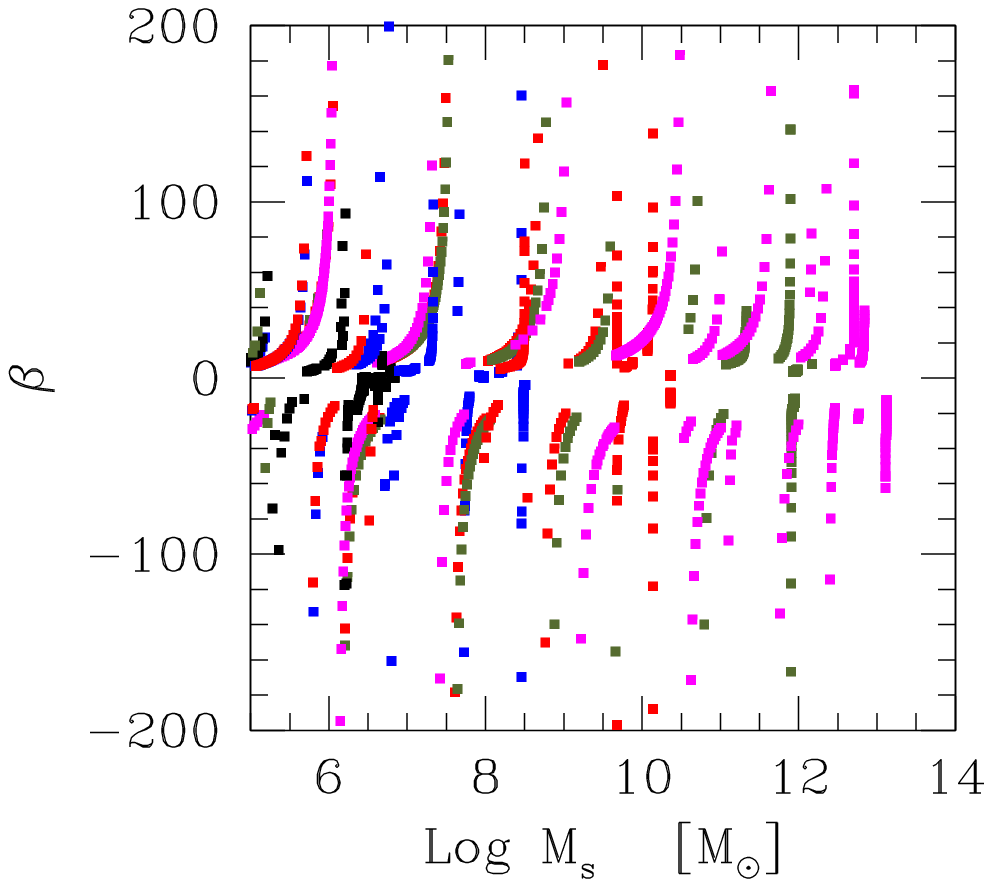} }
   \caption{ {\bf Left Panel}: The $\beta$ - $I_e$ relationship for models of the $N_k$=10 group with 
   $\tilde M= 10^6$ (black dots), $10^8$ (blue dots), $10^{10}$ (red dots), $10^{12}$ (dark olive green dots), and $10^{13}$ (magenta dots), $M_\odot$. {\bf Right Panel}: the $\beta$ - $M_s$ relationship for  models of the two groups $N_k$=10 and $N_k$=20,  the same masses, and the same color code as in the left panel. }
              \label{Fig:beta_lop_ie_rel}
    \end{figure*}

In a series of papers,  \citet{Donofrioetal2017,Donofrioetal2019,Donofrioetal2020, DonofrioChiosi2021,Donofrio_Chiosi_2022,Donofrio_Chiosi_2023a,Donofrio_Chiosi_2023b, Donofrio_Chiosi_2024, Donofrio_Chiosi_Brevi_2025} proposed a new way of reading the diagnostic planes, projections of the FP of galaxies, in terms of two parameters, $\beta(t)$ and $L'_0(t)$ of the \Lsigbtempo\ relation already presented in the introduction.
The parameters $L'_0(t)$ and $\beta(t)$ were found to describe the distribution of galaxies in the various diagnostic planes and to predict the path of a galaxy in any of these planes in the course of evolution, by the natural changes of the various physical quantities defining a galaxy in the FP.  We refer to it as the $\beta-L_0'$ theory, a quick summary of which is given in Appendix \ref{app3_beta_theory}. 
The key result is that the  parameters $L$, $M_s$, \re, \Ie\ and $\sigma$ of a galaxy fully determine its path on FP and its projections according to the information encoded in the parameters $\beta$ and $L'_0$.

\begin{figure}    
   \centering
   {
\includegraphics[scale=0.45]{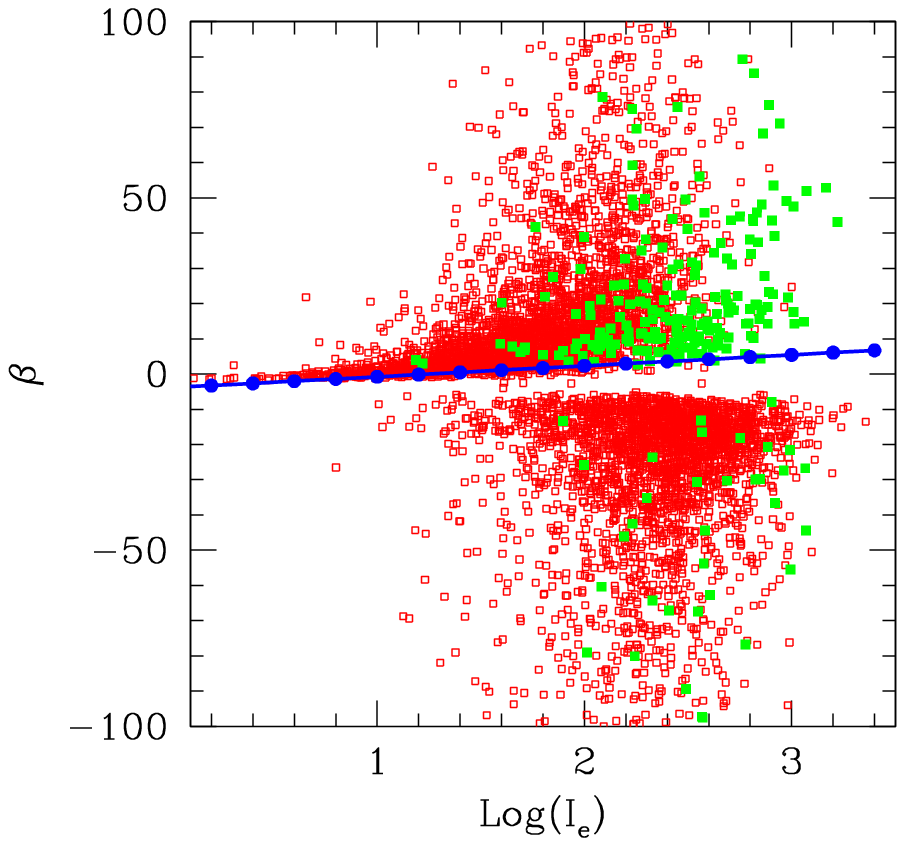} }
\caption{ The $I_e-\beta$ plane for the MANGA (open red squares) and WINGS data (green dots) as well as the  fully analytical merger models calculated by \citet{Donofrio_Chiosi_Brevi_2025} (blue dots along the black line) from whom the figure is taken however adapted to our data. }
              \label{Fig:Ie_beta_MA_WI}
    \end{figure}

Figure \ref{Fig:Ie_beta_MA_WI}  provides a comparison between the values of $\beta$ and $I_e$ for  MANGA and WINGS galaxies and the models of \citet{Donofrio_Chiosi_Brevi_2025} (solid line with dots). 
On the observational side,
the parameter $\beta$ is very low at very low surface brightness and progressively acquires large values (both negative and positive) for the high surface brightness galaxies, thus showing a large scatter.
In contrast, all the models are squeezed along a straight line which runs along the border of the distribution of real galaxies in the positive semi-plane. 
It is worth recalling here that those models were calculated by means of the simple application of the virial theorem and iterative  application of eqn.
(\ref{Var_Ref_Gal_a}) starting from seed values of $I_e$, $R_e$, $M_s$, $L$ and $\sigma$ taken from a infall model at a certain value of the age (redshift).
Similar result was found by \citet{Donofrio_Chiosi_2023a} in the case of the simple galaxy models with infall and no inclusion of any simulation of mergers.   
In contrast, the theoretical simulation of Illustris-TNG  and the previous ones of Illustris show similar dispersion in the $\beta-I_e$ plane. It goes without saying that the Illustris models framed and evolved into the hierarchical scenario  rest onto the merger mechanism by construction. Nevertheless, the physical reasons for this intriguing situation have never been clarified.  

\citet{Donofrio_Chiosi_Brevi_2025} carefully analyzed this issue and reached the conclusion  that there is still something missing in their scenario and that the use the Virial Theorem per se    could not  solve the problem, that is it can not account for the large dispersion of the data. The arguments brought  by \citet{Donofrio_Chiosi_Brevi_2025} to explain the reasons of the discrepancy can be summarized as follows: 
 (i) First they noted that  the gap of triangular shape void of galaxies falling in between the distribution on the positive and negative semi-planes corresponds in the $I_e$-$R_e$ plane to objects of relatively low surface brightness $I_e$ (and also luminosities) and intermediate radii $R_e$ that do not exist in our observational sample. They should correspond to the so-called diffuse galaxies that are difficult to identify (and perhaps are rare in the sky). So the gap is not a problem and can be left aside. 
 (ii) Second, they argued that several uncertainties affecting the estimates of luminosity, radii, masses etc. could blur the distribution of galaxies in the $I_e-R_e$ plane and consequently affect the evaluation of $\beta$ and $L'_0$. A numerical experiment in which the values of $I_e$, $R_e$, and $M_s$ were randomly varied (within the uncertainty affecting these observational data, that is 20\%) showed that a large scatter of $\beta$ with positive and negative values was possible. 
(iii) Another source of uncertainty taken into consideration was  the parameter $k_v$, empirical   evaluation of the morphology of the galaxy, which is unknown in most cases,  at least in the samples under examination.
Catching the right combination of all these parameters is highly unlikely. 
(iv) They also argued that from a mathematical point of view, the very large  both positive and negative values of $\beta$ (and $L'_0$) reached by some galaxies could reside  in  the definition of $L'_0$ and $\beta$. In brief,
relations (\ref{eq2}) of Appendix \ref{app3_beta_theory} show that $\beta$ depends on $log L'_0$ which in turn contains the quantity ($1- 2A'/A$) at the denominator, where $A'$ is $2log\sigma$ while the $A$, expressed by means of $I_e$, $G/k_v$, $M_s$, $L$ and $R_e$, can also be reduced to $2log\sigma$. Therefore,  $\beta$ should diverge when the denominator  tends to zero.  So high values of $\beta$ and $L'_0$ should be the signature that the two derivation of $\sigma$ are in mutual agreement. This implies that only the right combination of $I_e$, $G/k_v$, $M_s$, $L$ and $R_e$ could yield this result. This could be a viable explanation of the large scatter in $\beta$ and $L'_0$.  However the questions soon arise: is this a frequent or rare event in a galaxy? What is the deep cause of it?

\begin{figure}    
   \centering
   {
   \includegraphics[scale=0.45]{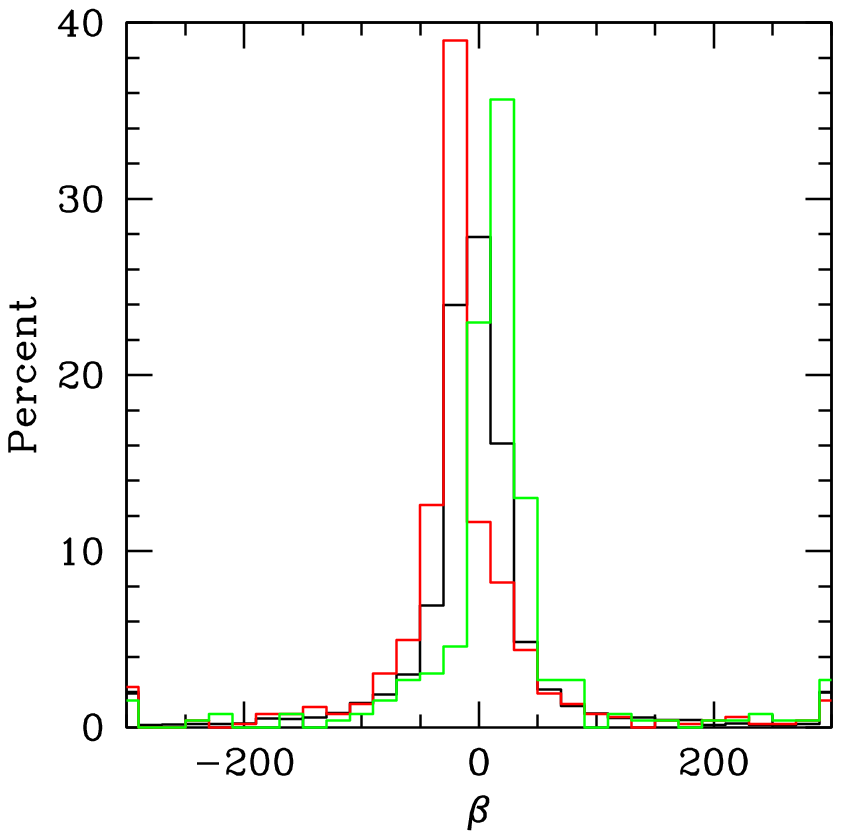} }
   \caption{ Histogram of the percentage  distributions of $\beta$'s of the  MANGA and WINGS galaxies. MANGA: the black is for  all galaxies (6291 objects) and the red line for E galaxies 414 objects). WINGS:  the green line are the ETGs (261 objects). This figure is similar to that of \citet{Donofrio_Chiosi_Brevi_2025}.}
              \label{Fig:histo_beta}
    \end{figure}

To cast light on this issue they looked at the relative number of galaxies as a function of $ \beta$. The MANGA sample contains 6291 objects of all types, of which  414 are ETGs (only E galaxies, S0 are left aside) and 2075 LTGs (Sa+Sb+Sc lumped together).  The WINGS sample  contains 261 ETGs only.  The percentage  distribution of all these objects is shown as a function of  $\beta$ in steps of 10 in Fig.\ref{Fig:histo_beta}, see also \citet{Donofrio_Chiosi_Brevi_2025}. The histograms displays the three populations with different colors: (i) in the case of MANGA, black for all galaxies (6291 objects) and red for ETGs (414 objects), and finally 2075 objects for Sa+Sb+Sc not shown here; (ii) green in the case of WINGS with 261  ETGs. In MANGA the total population is centered at small values of 
$\beta$ ($ -10 < \beta < +10$), while ETGs  peak in between $ -10 < \beta < 0$). In brief  ETGs are more shifted to the negative domain while LTGs are more shifted towards the positive domain. In WINGS, ETGs peak at $0 < \beta <20$.  Finally, both ETGs and LTGs have tails towards high values of $|\beta|$ on both semi-planes. Assuming as tail all galaxies falling below $\beta=-40$ and above $\beta=40$,  the estimated total  fractions in the tails and peaks amount to 0.15 (left), 0.45 (peak), 0.27 (right)  for ETGs and 0.09 (left), 0.43 (peak) and 0.09 (right) for LTGs. 

Recalling that  the typical dynamical timescale for a galaxy is about 0.5 Gyr, this is also the timescale required to  lose memory of any dynamical perturbation like a merger. On the other hand, if the merger induces star formation accompanied by a variation in total luminosity, about 3 Gyr is the timescale required to lose memory of the luminosity perturbation (see the data in Fig.17 of \citet{Donofrio_Chiosi_Brevi_2025} ). It follows that while the memory of the dynamical event is quickly lost, that of the luminosity variations lasts longer. Therefore, all quantities that directly or indirectly derive from the luminosity keep memory of the event for longer time.  
What we learn from the study of \citet{Donofrio_Chiosi_Brevi_2025} and the present data  is that  
mergers and luminosity effects must be simultaneously taken into account   to explain the scatter   on the $\beta$-$I_e$ plane (observational data and theoretical models).

\begin{figure*}    
\centering
   {
   \includegraphics[scale=0.45]{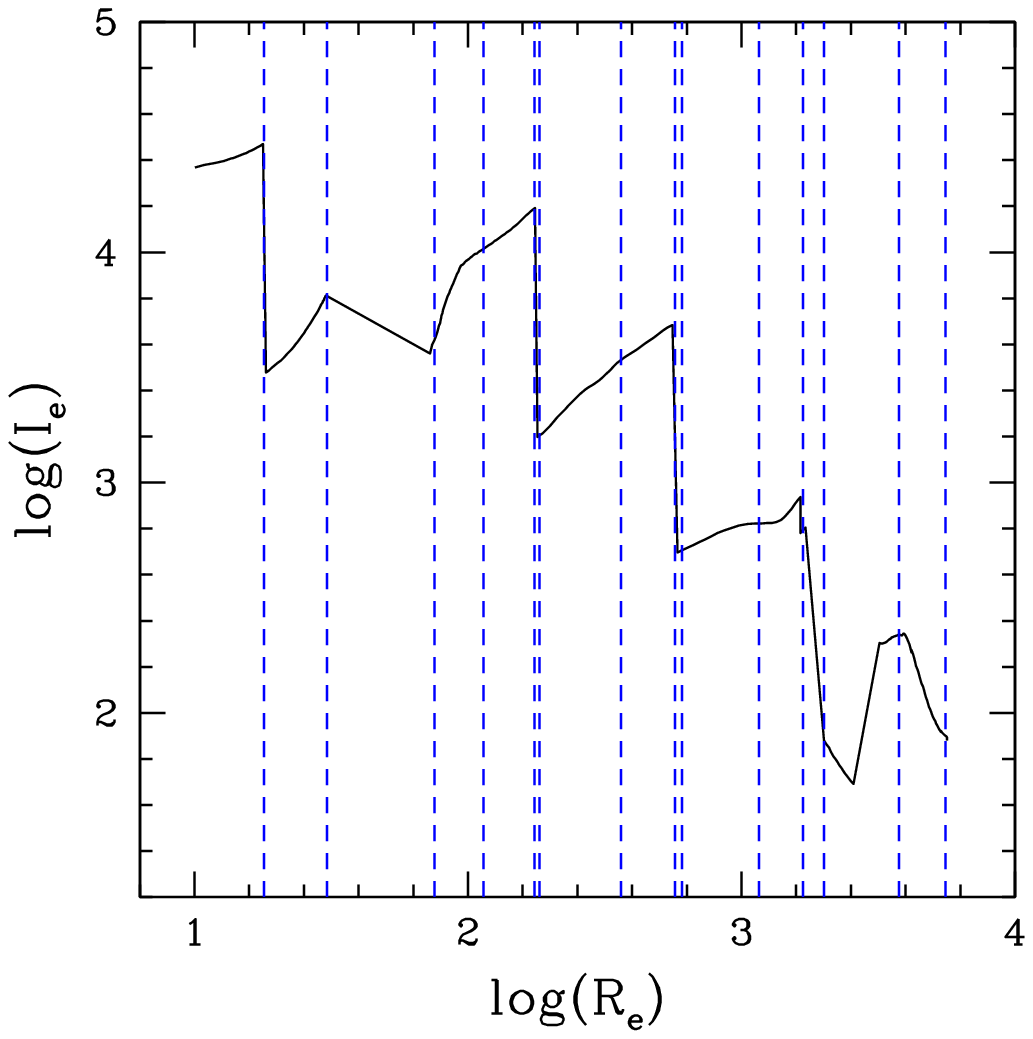}
   \includegraphics[scale=0.45]{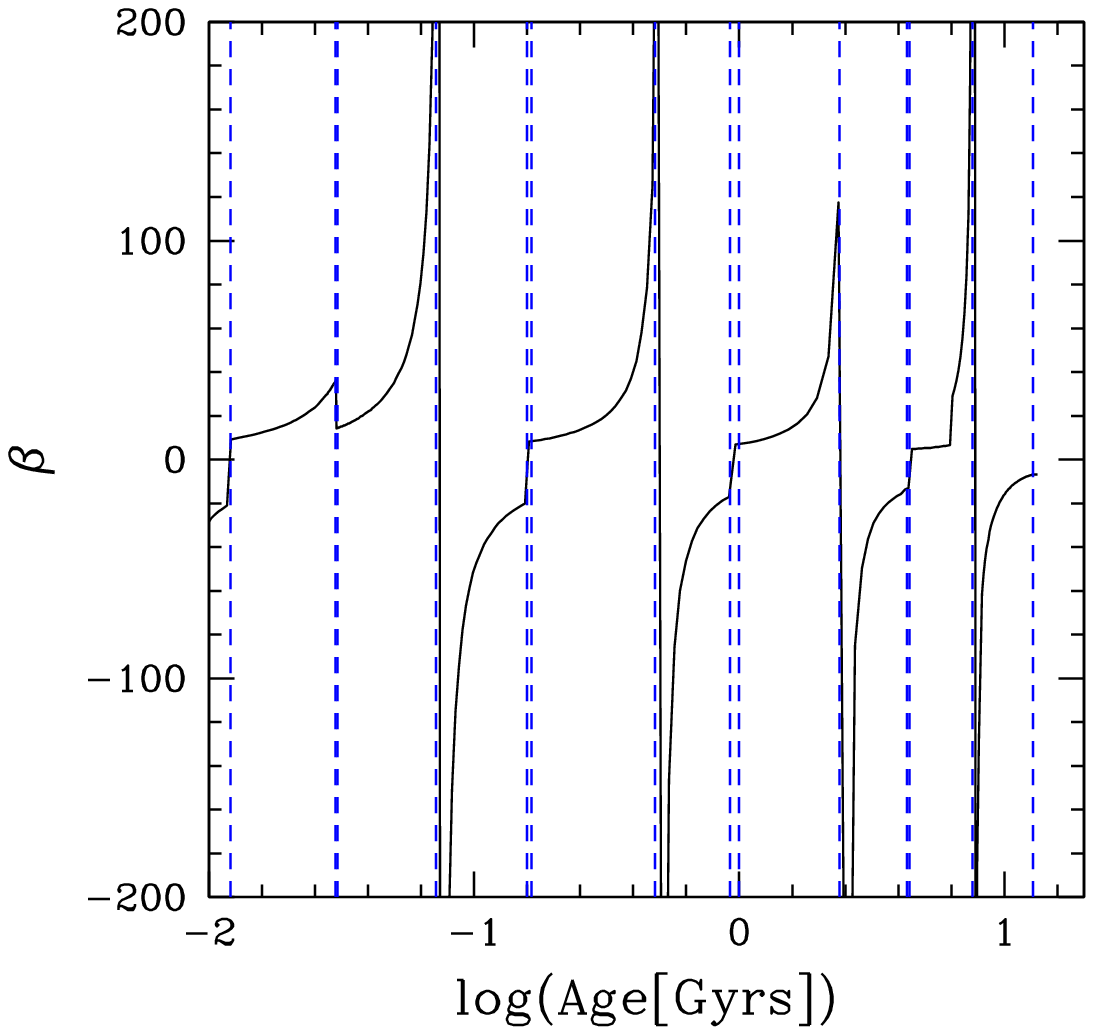}
   }  
\caption{  {\bf Left Panel}: the $I_e$- versus $R_e$ relation 
for the model with $\tilde M = 10^{10}\, M_{\odot}$ undergoing 10 mergers. The vertical dashed blue lines mark the  stages at which we note sudden variations of $I_e$.
{\bf Right Panel}:
$\beta$ versus age (Gyrs) for the same model shown in the left panel. The ages drawn by the vertical lines correspond to the stages indicated by the vertical lines in the left panel. }
 \label{Fig:beta_Age_ReIe}
\end{figure*}

This is not the end of the story, because the present simulations with  better and more realistic representation of a merger, show an aspect that had never been taken into consideration until now.
 We start from showing in Fig.\ref{Fig:beta_Age_ReIe} the relationships  $I_e$ ($L_{\odot} / pc^2$) versus $R_e$ (in kpc), left panel,  and  $\beta$ versus age (in Gyrs), right panel,  for the galaxy with $\tilde M = 10^{10}\, M_{\odot}$ undergoing 10 mergers. 
The vertical dashed lines show the ages at which the mergers take place. In this case,  there are  two major mergers  at about 4.3 and 6.2 Gyrs which cause important variations in the parameters $\beta$, $I_e$ and $R-e$ that are shown in the panels of Fig.\ref{Fig:beta_Age_ReIe} as indicated.

Commenting the panels of Figs \ref{Fig:beta_Age_ReIe} and the $M_s$-age and $R_e$-age relations of the $\tilde M = 10^{10}\, M_{\odot}$ case (much similar to those in \ref{Fig:Ms_Rs_Lb_sig_Age}) 
 we note what follows: (i) in the $\beta$-age relationship, the rise of $\beta$ starts in coincidence of the major sudden changes of the surface brightness Ie;   
ii) The relation $I_e$-$R_e$ is always of saw-tooth-type;  
iii) Looking at  ages at which mergers occur in the $R_e$-age relation and comparing these with the ages at which  minima/maxima of $I_e$ occur on the $I_e$-$R_e$ plane,   we note a strict coincidence;
(iv) In the $I_e$-age plane, we  see that the points selected in the $I_e$-$R_e$ plane coincide with the stage at which $I_e$ either increases or decreases;
(v) Finally, the ages of the peaks in $I_e$-$R_e$ plane coincide with those  noted in $\beta$-age plane.  The peaks of $\beta$ occur when the half-way stages in the variations of $I_e$ and $R_e$ are reached.

In a few words, when a merger occurs causing a large variation of $I_e$ and $R_e$, the parameter $\beta$ exhibits a large peak both positive and negative. At the start of the event, in correspondence to the rapid variation of $M_s$ and $R_e$, the variation of $\beta$ begins, and $\beta$ moves toward its peak value.  The peak can be  either positive or negative because the variations of $I_e$ and $R_e$ follow  a saw-tooth-like curve, that is the variation of $\beta$ passes two times through the critical stage. The mathematical source of the rapid twofold change (positive or negative) of $\beta$ resides the definition of  $\beta$ itself as already pointed out by \citet{Donofrio_Chiosi_Brevi_2025} however without identifying the real cause of it. The clue result is that high values of $\beta$  should be the signature of important variations of $I_e$ and $R_e$  caused by a recent merger. 

As last remark, we note that mergers are not the only cause of changes in $I_e$. In reality there are others intrinsic to the stellar populations.  In brief, during the first Gyr of a galaxy's life  the first generations of stars may undergo two sudden variations in the integrated luminosity (hence $I_e$)  caused by the appearance of the first AGB stars at about 0.1 Gyr and of the first RGB stars at about   0.8 Gyr in the mix of stellar populations generated by star formation\footnote{The effect of AGB and RGB stars on the integrated light and colors of stellar population was thoroughly investigated by \citet{Chiosi_etal_1988} in the case LMC star clusters.}. In the infall models with the mass accretion time scale $\tau= 1$ Gyr we are using, within the first Gyr or so in addition to quick growths  of $M_s$, $R_e$ and $L_s$ there are also two rapid variations of $L_s$ caused by the appearance of AGB and RGB stars in the first stellar populations. Therefore, we expect to see two discontinuities in the variation of $\beta$ at ages  within the first Gyr.
Later this effect will quench off due to the decreasing effect of AGB and RGB  stars on the total mix of stellar populations built up in the galaxy. Therefore, in absence of mergers, the initial fluctuations of $\beta$ caused by the stellar populations soon disappear. In contrast, it may show up again  a number of times in presence of repeated mergers (one fluctuation per merger). 

The main conclusion of this discussion is that mergers by inducing relevant variations of the luminosity (these latter are even enhanced if new star formation occurs) are the primary cause of the large positive/negative scatter of $\beta$. Thanks to it, the finger-prints of a  merger may last for long time. According to the present quick analysis about 20\% of the total population of galaxies in the nearby Universe could be in  post-merger conditions.

\section{Scale relations at high redshifts}

In this section we examine the behavior of our models at high redshift and compare them with the large scale numerical simulations of galaxy formation and evolution in the hierarchical scenario of the Illustris-TNG100 archive \citep[][and references therein]{Vogelsberger_2014a, Vogelsberger_2014b}.
We examine four different ScRs at three different values of redshift, namely  the \IeRe\ plane, $R_e$-$M_s$ plane, and the \Lsig\ plane at the z=4, 2, and 0 (the reference case). We use the $N_k$=10 group, and three different model galaxies with $\tilde M = 10^8, 10^{10}, {\rm and} 10^{12} \, M_\odot$.

\begin{figure*}    
   \centering
   {\includegraphics[scale=0.40]{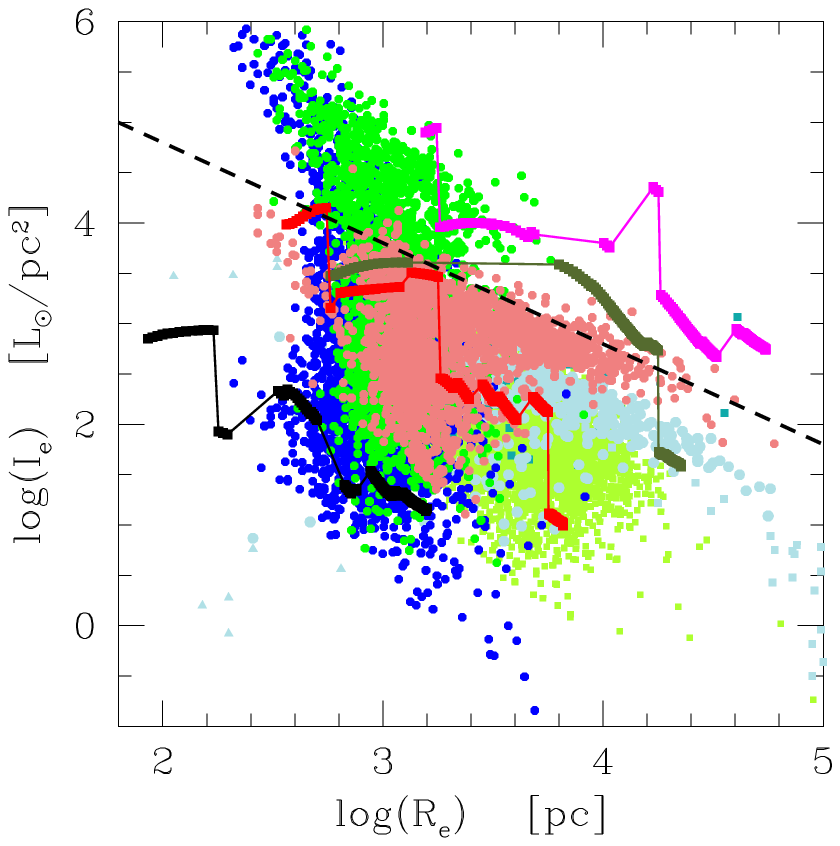} 
    \includegraphics[scale=0.40]{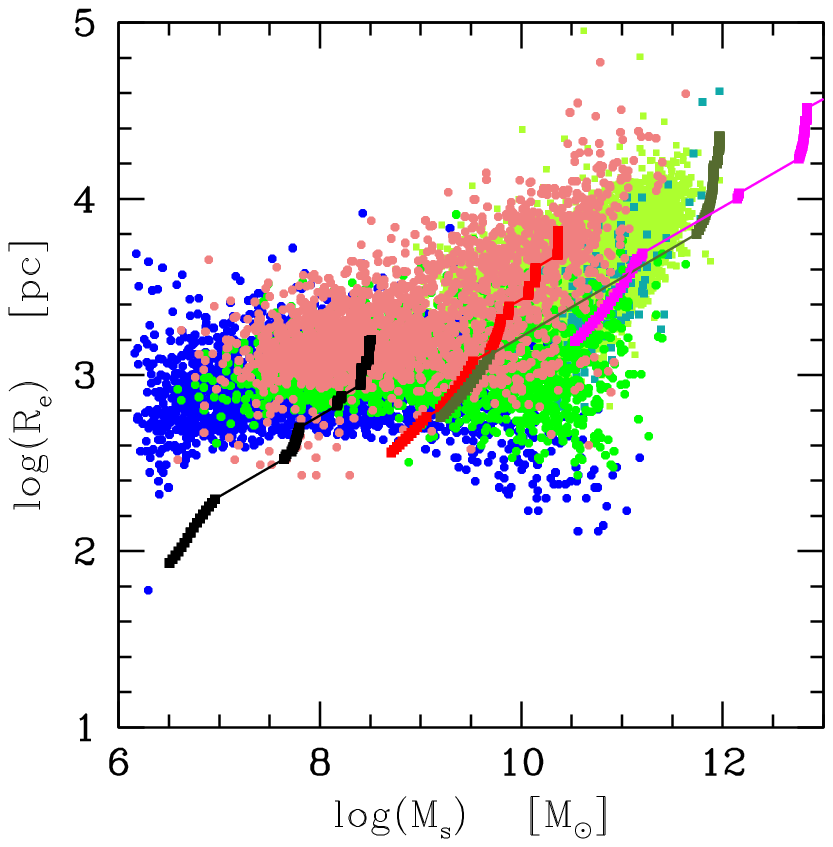}  }
   \caption{ $I_e$-$R_e$ plane (Left Panel) and $R_e$-$M_s$ plane (Right Panel) of data and models. The data are from the MANGA and WINGS samples lumped together (green dots). Theoretical models are from the Illustris-TNG100 large scale cosmological simulations at different redshift: z=4 (red dots), z=2 (blue dots), and z=0 (yellow dots). The solid lines are our semi-analytical models of the $N_k$=10 group for $\tilde M = 10^8\, M_\odot$  (black line and dots), $10^{10}\, M_\odot$ (red line and dots), and $10^{12}\, M_\odot$ (magenta line and dots) along part of their evolutionary path from the first merger to the present time.  }
   \label{Fig:shift_IeRe_ReMs}
    \end{figure*}

The \IeRe\ plane, The left panel of Fig.\ref{Fig:shift_IeRe_ReMs} shows the \IeRe\ plane of the MANGA plus WINGS  databases lumped together  at z=0 (green dots) with superposed the  Illustris-TNG100 simulations at different redshift, that is  z=4 (red dots), z=2 (blue dots), and z=0 (yellow dots). The solid lines are our semi-analytical models of the $N_k$=10 group for $\tilde M = 10^8\, M_\odot$ (black lines), $10^{10}\, M_\odot$ (magenta lines), and $10^{12}\, M_\odot$ (brown lines) along part of their evolutionary path from z=10 (formation) to z=0 (present time) \footnote{We  recall that the half light radius provided by the MANGA data is somewhat different from the half mass radius given by simulations.}. 

\begin{figure*}    
   \centering
   {\includegraphics[scale=0.40]{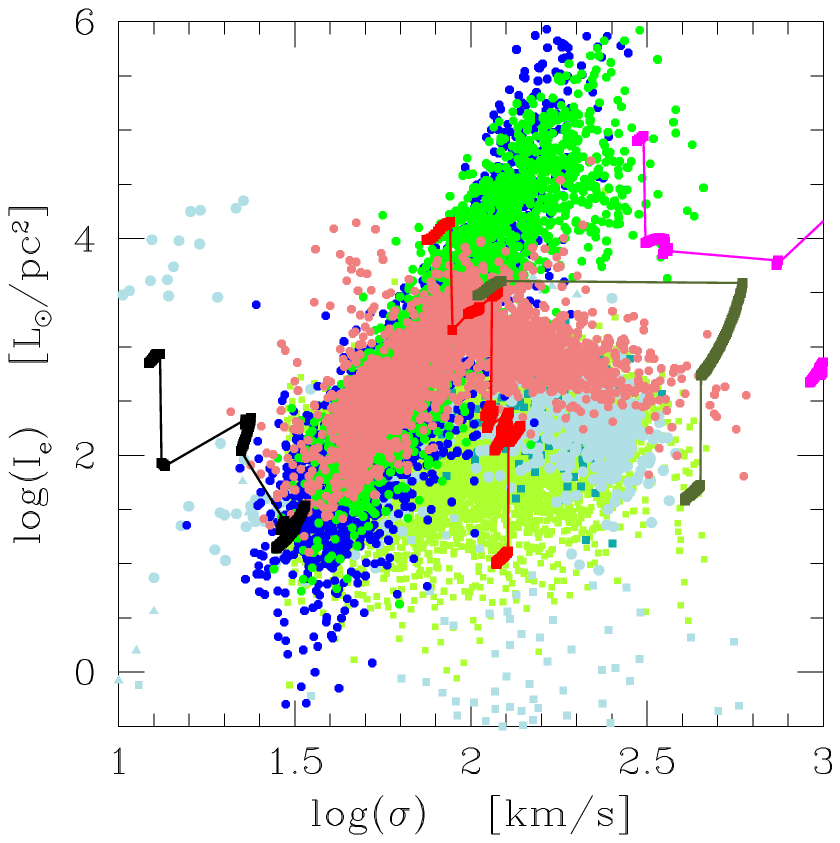} 
    \includegraphics[scale=0.40]{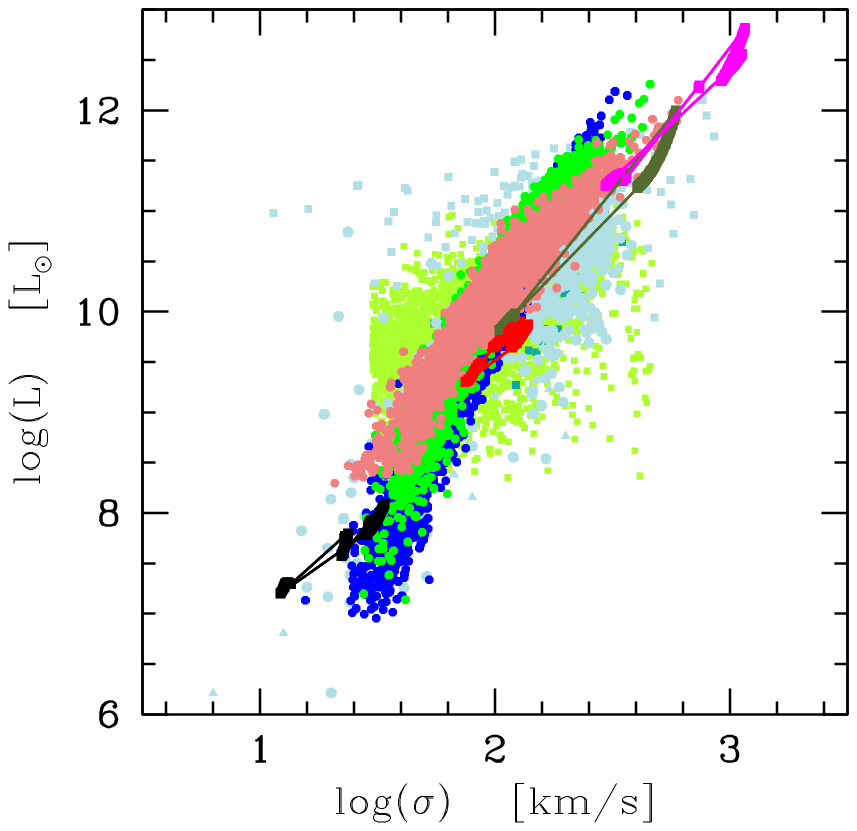}  }
   \caption{ The  $Ie_e$ - $\sigma $ relation (left panel) and the $L$ -$\sigma$ relation (right panel) of the same data, models, and color code as in Fig.\ref{Fig:shift_IeRe_ReMs}. }
   \label{Fig:shift_IeSig_LSig}
    \end{figure*}

In this plane  we note that the observational data and the  Illustris-TNG100 models at z=0 overlap and all of the them fall below the classical ZoE line.  Going to higher redshift (z=2 and 4) the distribution of the galaxy models becomes steeper    and a large number of them falls above the ZoE. 
Looking at the comparison in more detail, it is clear that the simulations reproduce many of the features observed in the \IeRe\ plane. For example, the $\Lambda$-shaped distribution of galaxies observed in this plane, in which the more massive galaxies are distributed along a tail  with a slope $\sim-1$, while the dwarf galaxies spread out on a cloud that extends along the  vertical direction.  It is worth recalling, however, that in the case of the MANGA data, the presence of the spiral galaxies in the sample produces a more uniform distribution so that   the $\Lambda$-shaped feature  is less visible.
The lack of objects at low radii and very low surface brightness, that appears as a boundary, is due to selection effects. Only galaxies above a given luminosity are present in the observational data as well as in simulations.
Finally, we confirm the  existence of the ZoE for the z=0 data with a slope close to $-1$, a value predicted by the VT, and the lack of galaxies beyond this limit (at least at z=0). The origin of this boundary is still a matter of debate.  However looking at the data for higher redshift, the classical ZoE is amply trespassed by many galaxies.  The   
suggestion arises that there is a new ZoE represented by the relation $\log Ie = -2.4 \log Re + 6.0$ holding good from  z$=0$ to z$\simeq=4$ (the classical one established only on the z=0 data is $\log Ie = -1.0 \log Re + 6.5$).  It is also amazing to note that the new relation extended to the region of clusters of galaxies (see Fig. \ref{Fig:Ie_Re_plane} in which the galaxy clusters of \citet{Burstein_etal_1997} are displayed) can account may also solve an old discrepancy. More precisely, the ZoE established for galaxies, if extended to regions of galaxy clusters, runs somewhat above these objects or vice-versa the galaxy clusters fall always below the classical ZoE. The reason for all this is not known. The new ZoE based on theoretical simulations, which  includes high redshift objects and is steeper than the classical ZoE by about  a factor of 2.5, does still agree with ETGs and is also the boundary of galaxy clusters.  Finally, the solid lines of different colors show the path of our semi-analytical models of the $N_k$=10 group with mass $\tilde M = 10^8\, M_\odot$  (black lines), $10^{10}\, M_\odot$ (magenta lines), and $10^{12}\, M_\odot$ (brown lines). Remarkably, in their early stages at redshift greater than 1 they fall into the regions drawn by the reference models of the Illustris-TNG100 archive. In conclusion, the picture emerging from comparing  different sets of models based on different hypotheses and calculations is fully self-consistent.

The $R_e$-$M_s$ plane. The distribution of galaxies in  this plane at varying the redshift is shown in the right panel of Fig.\ref{Fig:shift_IeRe_ReMs}. 
Once more, simulations are able to reproduce the curvature of the distribution at  high masses and the zone of exclusion on the right side with slope $\sim 1$. Only the scatter in radius is not fully matched. The observational data gives a much larger distribution of \re.
Most likely, this happens because spiral galaxies are added to the distribution.
The simulations also predict  radii of the low mass galaxies (those with $M_s<10^9 M_\odot$) that are not present in our observational database. The flattening of the distribution at low mass galaxies  is clearly visible both in the observational data and theoretical simulations. In contrast, the tail formed by the massive galaxies is well matched by the models.
We note that the limit of $M_s$ at low mass hand does gradually shifts toward higher masses at decreasing redshift,  and the same time at increasing mass the upper limit of $R_e$ shifts toward larger radii.  This is partly due to different low mass resolution limits in the simulation and partly is due to the physical facts that galaxies tend to increases their mass and hence radius as time goes by accumulating more and more mass by infall and/or mergers. The semi-analytical models of different mass $\tilde M$ follow this trend. The upper boundary obeys the relationship  $\log R_e  = 0.24 \log M_s + 1.57$.

The $I_e$-$\sigma$ plane. This relation is shown in the left of Fig.\ref{Fig:shift_IeSig_LSig} for the same models  and color code as in Fig.\ref{Fig:shift_IeRe_ReMs}. Also in this case, the slope of the relationships drawn by the Illustris-TNG100 models varies with the redshift. It becomes flatter and flatter at decreasing redshift. Our semi-analytical models are compatible with the different distributions of the Illustris-TNG100 models at increasing age (decreasing redshift). 

The \Lsig\ plane. The right panel of Fig. \ref{Fig:shift_IeSig_LSig} shows the \Lsig\ relationship at different redshift. It is clear that in this case  there are some difficulties in reproducing the slope and the large scatter of the data. At high redshift the Illustris-TNG100 draw a  a locus that is too steep compared to the part of the path followed by the semi-analytical model in their early evolutionary stages. Amazingly enough the late stage of the three sequences on display mimic the distribution of the Illustris-TNG100 data. In contrast the mean slope of the semi-analytical models is in nice agreement with the observational data and the Illustris-TNG100 models at z=0, but cannot explain the large scatter.  
Furthermore, the slope of the theoretical relation is slightly higher with respect to that of the data. Observations indicate that the galaxies of low luminosity progressively depart from the \Lsig\ relation and are distributed along a horizontal cloud of points. On the other hand, simulations predict a \Lsig\ relation holding  over the whole range of magnitudes with nearly constant  scatter everywhere.

A number of explanations can be found. It should be recalled for example that the MANGA data  derive $\sigma$ from the IFU. It follows that the velocity dispersion depends on the aperture of the fiber elements with respect to the galaxy dimension and the orientation of the galaxy. Furthermore the MANGA data contains also spiral galaxies whereas the \Lsig\ was  originally drawn  only for pressure supported systems like E and S0 galaxies  so that the disagreement could be due to the different definition of velocity dispersion between early and late type of galaxies. 
On the theoretical side, simulations provide the mean velocity dispersion of the star particles within the half mass radius. The definition of half mass radius is far from being univocal for galaxies of different type. We conclude by saying that
it is not the aim of this work to discuss the possible explanations for the observed differences between observations and simulations. The qualitative comparison made here wants only to demonstrate that the present hydrodynamic simulations are able to reproduce many of the main 
features of the observed distributions in all the ScRs generated by the FP projections.

\section{Conclusions}\label{sec:10_concl}

In this study, first we aimed to propose and validate a simple semi-analytical model of galaxy formation in the hierarchical scheme as a proxy of real galaxies and  theoretical models obtained by means of the large scale cosmo-hydro-dynamical simulations; second to analyze the various projection  planes of the Fundamental Plane of galaxies at the light of the theoretical results to our disposal (observational data, large scale simulations, and  simple analytical model)  in a unitary conceptual framework.
Main results  are:

(i) The model we propose is a simple yet effective tool to  highlight the  role played by mergers in shaping the relations among the various physical variables characterizing  a galaxy that are indicated  by the observational data, e.g. the relationship between  the luminosity and velocity dispersion, between the  radius and mass, between  specific intensity and  radius and others. In other words, to interpret the different projections of the Fundamental Plane of the galaxy populations from  the multi-dimensional space of their physical variables onto distinct two dimensional planes. 
The achievement of this study is the development of a semi-analytical model of mergers  that does not require extensive numerical N-Body hydrodynamic simulations. 

(ii) The present model improves and supersedes the previous version of   
 \citet{Donofrio_Chiosi_Brevi_2025}  by including a simple yet realistic description of the way galaxy form, evolve, and undergo repeated mergers during their lifetime. 

(iii) The results of this model nicely match the basic Scale Relations exhibited  by galaxies of different type, mass and size and  mass and at the same time they provide a reasonable interpretation of the large scatter seen in the same Scale Relations.
 
(iv) The success of the model resides in the slide ruler algorithm, the adoption of infall models as building blocks of a merger, and the use of the virial theorem to  derive the energy injection and sharing among the two merging galaxies.  The Virial Theorem ultimately determines  the velocity dispersion and radius at each time step of the evolution of a galaxy. The formalism in use is simple and leads to meaningful  results at no cost in terms of computing time and resources. 

(v) A large number of simulations are made at varying the parameters, and with the aid of these, the usual diagnostic planes are derived and interpreted.  The results are found to be fully compatible with both observational data and extensive hydrodynamic simulations.  We limit ourselves to highlight here that the our simple semi-analytical models of mergers provides a simple explanation for the large dispersion shown by the data in the diagnostic planes and  also cast light on some unexpected sub-trends noticed in the $L-M_s$, $L-\sigma$ and $M_s-\sigma$ planes.   At present, the model cannot deal with  merger induced star formation and also has some limitations in the possible combinations of galaxy masses and ages taking part to a merger. Work is in progress to cope with it. Finally, the inner structure of galaxies, ultimately determining their morphology, is simply out of reach. 

To conclude and avoid any misunderstanding, we clarify that our semi-analytical description of mergers does not intend to compete with those from detailed large-scale numerical simulations. It is only a simple-minded tool to  quickly investigate  the physical causes by which galaxies populate the diagnostic planes in the way they do. 
Thanks to its simplicity, our approach offers some advantages   with respect to  extensive numerical simulations. However, it cannot  yield the precise predictions  of these latter.

\begin{acknowledgements}
      M.D. thank the Department of Physics and Astronomy of the Padua University for the financial support. C.C. thanks the Department of Physics and Astronomy of the Padua University for the hospitality. 
\end{acknowledgements}

%
   \bibliographystyle{aa} 
   \bibliography{ScRs.bib} 
%



\begin{appendix}

\section{Fully analytical  models of galaxy  mergers} \label{app1_ana_merger}
 
In this section we present the  simple analytical model by \citet{Donofrio_Chiosi_Brevi_2025} to describe and predict the growth in radius as a consequence of the growth in mass by mergers together with the variations in luminosity, surface brightness, and velocity dispersion. 

The model strictly stands on the formalism  developed 
by \citet{Naab_etal_2009} for the case of a single object in virial equilibrium on which other objects of assigned total mass and energies fall in and merge together. Assuming energy conservation, the new system is also supposed to reach the virial equilibrium. The model of \citet{Donofrio_Chiosi_Brevi_2025} generalizes the formalism to the situation in which many successive mergers can occur. In addition to mass and radius the new model also considers the variations in luminosity and specific intensity.

\textsf{The Naab et al. (2009) case.} Let us start with the initial system characterized by mass $M_i$, half-mass radius $R_{e,i}$, velocity dispersion $\sigma_i$, luminosity $L_i$, specific intensity $I_{e,i}$. The kinetic ($K_i$),  gravitational ($W_i$),  and total ($E_i$) energies of the  initial  object are 
 
 $$ K_i = \frac{1}{2} M_i \sigma_i^2, \quad  W_i = - \frac{G M_i ^2}{ R_{e,i}}, \quad  {\rm and} \quad  E_i = K_i + W_i$$
 
 \noindent
 The virial equilibrium is expressed by $2 K_i + W_i = 0$ from which one derives the two identities  
 
 $$ K_i = - \frac{W_i}{2}  \quad {\rm and} \quad  E_i =  \frac{W_i}{2}. $$
 
 \noindent
 Suppose now that the system {\it i}  merges  another system {\it a} with  parameters $M_a$, $R_{a,i}$, $\sigma_a$, $L_a$,  $I_{a,i}$, $K_a$,  $W_a$,  and $E_a$, and that the virial condition is rapidly re-obtained after the merging. Finally, introduce the ratios  $\eta= M_a/M_i$ and $\epsilon = \sigma_a^2 /\sigma_i^2$. The total energy of the system is now $E_f = E_i + E_a$. After the merger event, the composite system will soon recover the virial equilibrium so that we can write
 
 \begin{equation}
  E_f = - \frac{1}{2} M_i \sigma_i^2 (1 + \epsilon \eta) = - \frac{1}{2} M_f \sigma_f^2 
 \end{equation}

 \noindent
 where $M_f = M_i + M_a = (1+ \eta) M_i$. It follows from it that 
 
 \begin{eqnarray} 
  \frac{M_f}{M_i}                &=& (1+ \eta)  \nonumber \\
  \frac{\sigma_f^2}{\sigma_i^2}  &=& \frac{1+ \eta \epsilon}{1+\eta}  \nonumber\\
  \frac{R_{e,f}}{R_{e,i}}        &=& \frac{(1+\eta)^2}{1+ \epsilon \eta} \label{A_single}  \\
  \frac{\rho_{f}}{\rho{i}}       &=& \frac{(1+\eta \epsilon )^3}{(1+ \eta)^5} \nonumber 
 \end{eqnarray}

 \noindent
 Depending on the parameters $\eta$ and $\epsilon$ the increase of the radius is different. If the mass is simply doubled $M_f = 2 M_i$,  $\eta =1$ and hence $\epsilon = 1$, the radius $R_{e,f} = 2 R_{e,i}$, and the velocity dispersion $\sigma_f = \sigma_i$.
  
\textsf{Extension to several mergers}. The formalism can be easily extended to the case of many successive  different mergers. 
For the sake of simplicity, we start considering an initial object characterized by the parameters $M_i$, $R_{e,i}$, $\sigma_i$, $L_i$, $I_{e,i}$ in suitable units and capturing other objects  characterized by the parameters $M_a$, $R_{e,a}$, $\sigma_a$, $L_a$ and $I_{e,a}$ in the same units. In this step of the reasoning, the luminosity and specific intensity are not included. The number of mergers is given by $N$ (integer). The captured objects are all the same and are small compared to the initial object.  Define also the ratios $\eta = M_a/M_i$  and $\epsilon = \sigma_a^2 / \sigma i ^2$. The final total mass is $M_f = M_i + N M_a$. The initial galaxy, the captured object, and the final object are all in virial equilibrium. Following the reasoning of the two-body case, we arrive to the expressions
 
 \begin{eqnarray}
 E_f &=& - \frac{1}{2} M_i \sigma_i^2 (1 + N \epsilon \eta)  \\
  E_f &=& - \frac{1}{2} M_f \sigma_f^2   
  \end{eqnarray}
 
 \noindent
 and the following scaling laws for the mass $M_f$, $\sigma_f^2$, $R_{e,f}$, and $\rho_f$ as functions of the corresponding values of the initial galaxy
 
 \begin{eqnarray}
 \frac{M_f}{M_i} &=& (1 + N \eta) \nonumber \\
  \frac{\sigma_f^2}{\sigma_i^2} &=& \frac{(1+N \eta \epsilon)}{(1+ N \eta)} \nonumber \\
  \frac{R_{e,f}}{R_{e,i}}  &=& \frac{(1+N \eta)^2}{(1+ N \epsilon \eta)} \label{A_many}  \\
  \frac{\rho_{f}}{\rho{i}}  &=& \frac{(1+N \eta \epsilon )^3}{(1+ N \eta)^5} \nonumber \\ 
  \end{eqnarray}
  
\textsf{Luminosity and specific Intensity}. To include in the scaling laws those for the luminosity and specific intensity and derive the total luminosity $L_f$ and the total specific intensity $I_{e,f}$ of the remnant galaxy, we need to check in advance  some preliminary aspects concerning the luminosity because of its complex nature, more complex than the case of the mass, radius, and velocity dispersion.  
The subject has been examined  by \citet{Donofrio_Chiosi_Brevi_2025} to whom the reader should refer. They arrived to

\begin{equation}
\frac {L_f}{L_i} = \frac {M_f}{M_i} = (1+ N \eta \theta) 
\end{equation}

\noindent
where $\theta$ is a parameter similar to $\eta$ and/or $\epsilon$ that can be estimated from the theory of population synthesis \citep[see][for more details]{Donofrio_Chiosi_Brevi_2025}.
If one is not interested to highlight the time dependence of the luminosity variation, the factor $\theta$ can be neglected by simply posing $\theta=1$.

\noindent
Since the specific intensity  is  given by $I_{e} = L /(2 \pi R_{e}^2)$,  the associated scaling relation is

\begin{equation}
\frac{I_{e,f} }{I_{e,i}} = \frac{ (1 + N \eta \theta) (1+ N\eta \epsilon )^2 }{(1+ N \eta)^4} .
\label{intensity_eq}
\end{equation}

\noindent
The last two relations complete the systems of equations (\ref{A_single})  and (\ref{A_many}).

\textsf{Including the kinetic energy of the merging object.}
For the sake of investigation, \citet{Donofrio_Chiosi_Brevi_2025}  also took into account that the captured object with mass $M_a$, velocity dispersion $\sigma_a$, also  carries the kinetic energy of the relative motion  $K_a = (1/2) M_a v_a^2$. Part of this energy is supposed to be absorbed by the receiving object and the rest be  dissipated in some way. Also in this case, the mass $M_a$  and velocity $v_a^2$ are parameterized in terms of the initial mass $M_i$ and initial velocity dispersion $\sigma_i$ 
by the relation $M_a= \eta M_i$ and  $v_a^2 =  \lambda \sigma_i^2$. After the merger, the composite system recovers the virial condition. With the aid of  some simple algebraic manipulations, we obtain

 \begin{eqnarray}
 \frac{M_f}{M_i} &=& (1+N \eta) \nonumber \\
  \frac{\sigma_f^2}{\sigma_i^2} &=& \frac{(1+N \eta (\epsilon +\lambda)}{(1+ N \eta)} \nonumber \\
  \frac{R_{e,f}}{R_{e,i}}  &=& \frac{(1+N \eta)^2}{1+ N \eta (\epsilon + \lambda}    \nonumber\\
  \frac{\rho_{e,f}}{\rho{e,i}}  &=& \frac{(1+N \eta (\epsilon + \lambda )^3}{(1+ N \eta)^5}  \label{B_many_lambda} \\
  \frac{L_f}{ L_i} &=& (1+ N \eta \theta)   \nonumber \\
  \frac{I_{e,f} }{I_{e,i}} &=& \frac{ (1 + N \eta \theta) (1+ N\eta (\epsilon+\lambda) )^2 }{(1+ N \eta)^4} \nonumber 
  \end{eqnarray}
 
 \noindent
 Note that for $\lambda=0$ the previous case is recovered.

\textsf{Random variations of $\eta, \epsilon, \lambda$, and $\theta$.}  These relations can be further generalized to take into account the possibility that the parameters $\eta$, $\epsilon$, $\lambda$, and $\theta$  can slightly vary from one galaxy to another and the merger event. The relationships of the system (\ref{B_many_lambda} become 

 \begin{eqnarray}
 \frac{M_f}{M_i} &=& (1+ \sum_j\eta_j  )  \nonumber \\
  \frac{\sigma_f^2}{\sigma_i^2} &=& \frac{(1+ \sum_j \eta_j \epsilon_j + \sum_j \eta_j \lambda_j)}{(1 + \sum_j \eta_j)} \nonumber \\
  \frac{R_{e,f}}{R_{e,i}}  &=& \frac{(1+\sum_j \eta_j)^2}{1+ \sum_j \eta_j \epsilon_j  + \sum_j \eta_j \lambda_j}  \nonumber \\
  \frac{\rho_{e,f}}{\rho{e,i}}  &=& \frac{(1+\sum_j \eta_j \epsilon_j + \sum \eta_j \lambda_j)^3}{(1+ \sum_j \eta_j)^5}  \label{A_many_rand}  \\
  \frac{L_f}{L_i} &=& (1+ \sum_j\eta_j \theta_j )  \nonumber\\
  \frac{I_{e,f} }{I_{e,i}} &=& \frac{ (1 + \sum_j \eta_j \theta_j ) (1+  \sum_j \eta_j \epsilon_j + \sum_j \eta_j \lambda_j)^2 }{(1+ \sum_j \eta_j)^4} \nonumber  
  \end{eqnarray}
 
 \noindent
where the suffix $j$ goes from 1 to $N$. In the summations  $\sum_j \eta_j$,  $\sum_j \epsilon_j$,  $\sum_j \eta_j \epsilon_j$, $\sum \eta_j \lambda_j$, $\sum \eta_j \theta_j$ the parameters $\eta_j$,  $\epsilon_j$,  $\lambda_j$, and $\theta_j$ are randomly  varied within suitable intervals by means of uniform distribution of random numbers between 0 and 1. Also in this case for $\lambda_j=0$ and  $\theta=1$ we recover the standard cases with no kinetic energy of the relative motion and no age effect on the luminosity.

\textsf{Changing the reference model step by step.} Before concluding this section, we like to present an alternative formulation of the problem which differs from the previous one from the initial hypothesis and yet leads to similar results. In the previous version the key hypothesis was: an initial object of mass $M_i$ that in all successive steps acquires a smaller object with  the same mass $M_a$, and energy smaller that those of the initial object. The mass $M_i$ and velocity dispersion $\sigma_i$ are always the reference values. 
  
In alternative we may suppose that at each merger the newly formed object becomes the reference system for the next episode. In other words, each merger is a two-galaxies event in which the parameters $\eta$, $\epsilon$ (and $\lambda$ and $\theta$) are referred to the last composed object.  These parameters must be continuously adjusted step by step to keep the perturbation small compared to the current object. This is achieved by simply dividing the parameters $\eta$, $\epsilon$, and  $\lambda$ by the current number of the step. The equations governing these models are 
  
 \begin{eqnarray}
  \frac{M_f}{M_i}               &=& (1 +  \eta) \nonumber \\
  \frac{\sigma_f^2}{\sigma_i^2} &=& \frac{ (1 + \eta (\epsilon+\lambda) )}{(1+  \eta)} \nonumber \\
  \frac{R_{e,f}}{R_{e,i}}       &=& \frac{(1 + \eta)^2}{ (1 +  \eta (\epsilon + \lambda) )}  \nonumber \\
  \frac{\rho_{f}}{\rho{e,i}}    &=& \frac{(1 + \eta (\epsilon + \lambda) )^3}{(1+  \eta)^5} \label{Var_Ref_Gal_b}  \\
  \frac{L_f} {L_i}              &=& (1  +  \eta \theta)   \nonumber \\
  \frac{I_{e,f} }{I_{e,i}}      &=& \frac{ (1 + \eta \theta) (1+ \eta (\epsilon+\lambda))^2 }{(1+  \eta)^4} \nonumber  
  \end{eqnarray}

Given an initial set of values for $M_s$,  $R_e$, $\sigma $, $L$,  $I_e$, a sequence of mergers at increasing mass is made by $N$ steps indicated by $J = 1,... N$. In order to keep the perturbation small compared to the ever increasing mass one  adopt the heuristic  prescription 
$\eta = \eta_r/J$, $\epsilon = \epsilon_r / J$, $\theta= \theta_r / J$  $\lambda= \lambda_r / J$,  where $\eta_r$, $\epsilon_r$, $\lambda_r$, and $\theta_r$ are the usual factors, randomly varied within suitable intervals.

\section{Short summary of  the infall models of galaxies }\label{app2_open_mod}

The infall models, proposed long ago by  \citet{Chiosi_1980} and ever since largely used in many papers on chemical evolution and photometric synthesis see for instance 
\citet[see for instance][]{Bressan_etal_1994,Tantaloetal1998,Chiosi_etal_2014}, were meant to simulate in a very simple fashion the formation and evolution of  galaxies  regardless of their  morphology (they are considered as point-mass entities that acquire gas and transform it into stars at a suitable rate). In brief, a galaxy forms by the collapse of primordial gas (baryonic mass $\rm M_B(t)$) in presence of a halo of dark matter ($M_D(t)$) and the transformation of ${M_{B}(t) }$ into stars, ${\rm M_{s}(t) }$. Dark matter plays a  secondary role by changing only the gravitational potential of the system. $\rm M_B(t)$)  is let increase with time according to

\begin{equation}
{  \frac { dM_{B}(t) }{dt} = {K_{B,0} } ~exp(-t/\tau) }   
\label{inf}
\end{equation}
\noindent
where $\tau$ is the accretion time scale and $K_{B,0}$ is a constant with dimensions [mass]/[time]. The proportionality  constant  $ K_{B,0}  $ 
is obtained from imposing that at the galaxy age 
${T_{G}}$ the value ${ M_{B}(T_{G})}$ is reached:

\begin{equation}
{   \dot{K_{B,0}} = \frac{M_{B}(T_{G})}{\tau 
[1 - exp(- T_{G}/\tau)]}   }  
\label{mdot}
\end{equation}

\noindent
Therefore, upon integration over time,  
${\rm M_{B}(t)}$ is given by 

\begin{equation}
{  M_{B}(t) =  \frac{ M_{B}(T_{G})}  { [1-exp (-T_{G}/\tau)] }  
                      {[1 - exp(-t/\tau)]  }  }
\label{mas-t}
\end{equation}

\noindent
Total baryonic mass $M_{B}(T_{G})$ is considered as one of the main parameters characterizing a galaxy.
Denoting with ${ X_{i}(t)}$ the current mass abundance of an element $i$ and
introducing the dimensionless variables 

\begin{equation}
{ G(t)=M_{g}(t)/M_{B}(T_{G})  }  \qquad {\rm and}  \qquad  { G_{i}(t)=G(t)X_{i}(t),   }      
\label{gas_fra_gas_i}
\end{equation}

\noindent
where by definition  ${ \Sigma_i X_i(t)=1}$.
The equations governing the time variation of the ${ G_{i}(t)}$ and hence 
${ X_{i}(t)}$ are

\begin{equation}
{   \frac {dG_{i}(t)} { dt}= - X_{i}(t) \Psi(t) + }
{  \int_{m_L}^{m_U} \Psi(t') 
\phi(M) Q_{i}(t') dm + { \frac{ d[G_{i}(t)]_{inf} } { dt }  }   }    
\label{degas_i}
\end{equation}

\noindent
where ${ \Psi(t)}$ 
is the rate of star formation in units of ${ M_B(T_G)}$, $t'=t - t_m$
with ${ t_m}$
the lifetime of a star of mass ${ m}$, ${ \Phi(m)}$  the initial mass 
function,
${ Q_{i}(t-t_{m})}$  the fraction of mass ejected by such a star in 
form of an
element $i$, and ${ \dot{[G_{i}(t)]}_{inf} }$  the rate of gas 
accretion. This is expressed by 

\begin{equation}
 \frac{d[G_{i}(t)]_{inf} }{dt} = \frac{X_{inf}{M_g(t) }}{dM(t)} { dt }  
\label{gas_inf}
\end{equation}
 
 All the details about  the 
${ Q_{i}(t-t_{m})}$ are
omitted here for the sake of brevity. They can be found in 
\citet{Chiosi_1980}, 
\citet{Matteucci_1991}, \citet{Tantaloetal1998} to whom the reader should refer. 

Key ingredients of the infall model of galaxy formation and evolution are: (i) The stellar birth rate, that is the number of stars of mass $m$ born in the 
interval ${ dt}$ and mass interval ${ dm}$, is given by
${ dN =\Psi(t,Z) \phi(m)~dm~dt }  
\label{birth}$, 
where ${ \Psi(t)}$ is the rate of star formation as a function of time and
enrichment, while ${ \phi(m)}$ 
is the initial mass function (IMF). 
(ii) Many formulations of the IMF can be found in literature. We adopt here, the classical law due to \citet{Salpeter1955} expressed by  $ {\rm \phi(m)=  m^{-x}}$  where ${\rm x=2.35}$. The IMF is
normalized by means of the parameter $ \zeta =  {\ rm \int_{m_*}^{m_u}\phi(m){\times}m{\times}dm} /{\int_{m_l}^{m_u}\phi(m){\times}m{\times}dm}  $
that is the fraction of total mass in the IMF above ${  m_{*}}$.  and 
the lower limit of integration ${ m_l}$. 
The upper limit of integration is fixed at ${ m_{u}=120 M_{\odot}}$, 
the maximum mass in
our data base of stellar models, while the mass limit  ${\rm m_*}$ 
is the minimum
mass contributing to the nucleo-synthetic enrichment of the interstellar medium
over a time scale comparable to the total lifetime of a galaxy. This mass is
approximately equal to ${\rm 1 M_{\odot}}$.   In our models  $\zeta=0.3$ that is a good choice in
because it yields   ${ M/L_{B}}$ ratios in agreement with the 
observational
data and is compatible with current prescriptions for
chemical yields,  
\citep[see][]{Tantalo_etal_1996, Tantaloetal1998, Chiosietal1998,Portinari_Chiosi_1999, Portinari_etal_1998}.

(iii) The rate of star formation is assumed to depend on the gas mass according to  ${ \Psi(t)= \nu M_{g}(t)^{k} }$   (otherwise known as the Schmidt law)    where $\nu$  and $k$ are two parameters. 
Finaly, the star formation rate, normalized to ${\rm M_B(T_G)}$, becomes $ {\rm \Psi(t)= \nu M_{B}(T_G)^{k-1} G(t)^{k}  }$.  
 
With the above law of star formation of equation, the time dependence of
the rate of star formation ${\rm \Psi(t)}$  depends on the type of model in usage.  In the infall model, owing to the competition between the rate of gas
infall and gas consumption by star formation, the rate of star formation starts
small, increases to a maximum and then declines. The age at which the peak
occurs approximately corresponds to the infall time scale $\tau$. 
The functional form that could mimic this behavior is the time delayed exponentially declining law:
\begin{equation}
 {  \Psi(t) \propto ({t}/{\tau})\exp\left(-{t}/{\tau}\right) }.
\end{equation}

The Schmidt law,  therefore, represents the link between gas accretion by infall and gas consumption  by star formation. 
iv) The exponent $k$ and efficiency $\nu$ (and its inverse the time sclae of star formation 
(${\rm t_{SF}=1/\nu}$ could be contrained to fit some observational property of galaxies. Here we simply assume $k=1$ and $\nu=1$  \citep[see][for a classical review]{Larson_1991}. This choice reduces the number of
parameters and appears fully appropriate to the present purposes.
Finally, the infall time scale $\tau$ could be related to the 
collapse time scale of the baryonic mass into the common potential well of BM +DM. For the sake of simplicity we adopt $\tau=1$ Gyr. 
(v) 
The infall models may also include the occurrence of galactic winds. When the energy deposit by
the supernova explosions and stellar winds from massive stars 
exceeds the gravitational binding energy of the gas, the galactic wind is let
occur halting further accretion and star formation, and ejecting all remaining
gas. The mass of the luminous matter is  frozen to the current value of the
mass in stars. This is the real mass of the remnant visible galaxy to be used
to compare theoretical results with observations. However, for the sake of
simplicity we will always refer  to the models by means of the asymptotic mass
${\rm M_{G}(T_{G})}$. 
The galactic wind  occurs when the total thermal energy of the gas equates the binding energy  of the gas. The thermal energy is determined by the energy input from  Type I and Type II supernovae and  stellar winds ejected by massive stars \citep[for all details, see][]{Tantaloetal1998}. Suffice to mention here that to evaluate the amount of total thermal energy acquired by the gas present in a galaxy and powering the ejection of galactic winds against the action of the gravitational field, one has to include cooling by radiative processes and to adopt a certain view of the relative distribution of BM and DM.  \citet{Bertin_etal_1992} and \citet{Saglia_etal_1992}  proposed a simple formulation of the problem by means of the ratios ${\rm R_{B}/R_{D}}$ and ${\rm M_{B}/M_{D}}$  where $R_B$ and $R_D$ are the radii of the baryonic and dark matter components  (assumed sphrerical symmetry). These ratios used to  derive the gravitational potential of the gas and are adopted in the models of \citet[][]{Tantaloetal1998}.     
vi) The age of galaxies  ${\rm T_{G}}$ depends on the the cosmological model of the Universe, e,g, the  
$\Lambda-CDM$ Universe with parameters  $H_0=71$  \kmsM,  $\Omega_m = 0.27$, and $\Omega_\Lambda = 0.73$. 

vii) The mass distribution of DM haloes in mind is according to the hierarchical scheme of 
\citet{Lukic_etal_2007}, \citet{Angulo_etal_2012}, and \citet{Behroozi_etal_2013} in which massive galaxies grow in number with respect to the low mass ones at decreasing redshift or increasing age of the universe.   Many other details of the infall models in use be found in 
\citet{Tantaloetal1998, Chiosi_Donofrio_Piovan_2023}. 

To conclude the key parameters identifying an infall models of galaxies are the timescale of mass accretion $\tau$ and the total mass  ${M_{B}(T_{G})}$.

\section{Short recap of the $\beta-L_0'$ theory}\label{app3_beta_theory}

In a series of papers,  \citet{Donofrioetal2017,Donofrioetal2019,Donofrioetal2020, DonofrioChiosi2021,Donofrio_Chiosi_2022,Donofrio_Chiosi_2023a,Donofrio_Chiosi_2023b, Donofrio_Chiosi_2024} proposed a new way of reading the diagnostic planes, projections of the FP of galaxies, in terms of two parameters, $\beta(t)$ and $L'_0(t)$ of eq. (\ref{eq1}), that is the \Lsigbtempo\ relation. These two parameters were found to describe the distribution of galaxies in the various diagnostic planes and to predict the path of a galaxy in any of these planes in the course of evolution, by the natural changes of the various physical quantities defining a galaxy in the FP. We refer to it as the I$\beta-L_0'$ theory.
In the following we drop the explicit notation of time dependence for the sake of  simplicity. The two equations representing the VT and the \Lsigb\ law are:

\begin{eqnarray}
 \sigma^2 &= & \frac{G}{k_v} \frac{M_s}{R_e}   \\
 \sigma^\beta &= & \frac{L}{L'_0} = \frac{2\pi I_e R^2_e}{L'_0}. 
\label{eqsig}
\end{eqnarray}

The first equation is the Virial Theorem itself, where $G$ is the gravitational constant, and $k_v$ a term that gives the degree of structural and dynamical non-homology. The presence of $k_v$ allows us to write $M_s$ (stellar mass) instead of the total mass $M_T$. $k_v$ is  a function of the S\'ersic index $n$, that is $k_v=((73.32/(10.465+(n-0.94)^2))+0.954))$ \citep[see][for all details]{Bertin_etal_1992, Donofrioetal2008}. In the two equations, all other symbols have their usual meaning. 
In these equations, $\beta$ and $L'_0$ are time-dependent parameters that depend on the peculiar history of each object. 

From these equations one can derive all the mutual relationships existing among the parameters $M_s$, $R_e$, $ L$, $I_e$, $\sigma$ characterizing a galaxy as a function of the $\beta$ parameter. Although they may be of some interest also in the context of this paper,  they are not repeated here. They can be found in \citet{Donofrio_Chiosi_2024}.
It suffices to recall that in all these relationships,  the slopes   depend  on $\beta$ while the proportionality coefficients, in addition to other structural parameters characterizing a galaxy, depend also on $L'_0$. This means that when a galaxy changes its luminosity $L$,  and velocity dispersion $\sigma$, and has a given value of  $\beta$ (either positive or negative), the effects of this change in the \Lsig\ plane are propagated in all other projections of the FP.  In these planes the galaxies cannot move in whatever directions, but are forced to move only along the directions (slopes) predicted by the $\beta$ parameter in the above equations. In this sense the $\beta$ parameter is the link between the FP space and the observed distributions in the FP projections.
In addition to this, it is possible to derive an equation holding for each single galaxy that looks like the classical FP. Combining eqs. (\ref{eqsig}), it is possible to write a FP-like equation involving the parameters $\beta$ and $L'_0$:

\begin{equation}
    \log R_e = a \log\sigma + b <\mu>_e + c
    \label{eqfege}
\end{equation}

\noindent
where  $<\mu_e>$ is the mean surface brightness $<I_e>$ expressed in magnitudes and the coefficients:

\begin{eqnarray}
a & = & (2+\beta)/3 \\ \nonumber
b & = & 0.26 \\ \nonumber
c & = & -10.0432+0.333(-\log (G/k_v) - \log (M/L) \\ \nonumber
  &   & -2\log (2\pi)-\log (L'_0)) \nonumber
\end{eqnarray}

\noindent
are written in terms of $\beta$ and $L'_0$. We note that this is the equation of a plane whose slope depends on $\beta$ and the zero-point on $L'_0$. The similarity with the classical equation for the Fundamental Plane  is clear. The novelty is that the FP is an equation derived from the fit of a distribution of real objects, while here each galaxy independently follows an equation formally identical to the classical FP, but of profoundly different physical meaning. In this case, since $\beta$ and $L'_0$ are time dependent, the equation represents the instantaneous plane on which a generic galaxy is located in the FP space and consequently in all its projections.

Finally, the equation system (\ref{eqsig}) allows us  to determine the values of $\beta$ and $L'_0$, the two basic evolutionary parameters. Let us  write the above equations in the following way:

\begin{eqnarray}
\beta [\log(I_e)+\log(G/k_v)+\log(M_s/L)+\log(2\pi)+\log(R_e)] + \\ \nonumber
    + 2\log(L'_0) - 2\log(2\pi) - 4\log(R_e) = 0 \\ 
 \beta\log(\sigma) + \log(L'_0) + 2\log(\sigma) + \log(k_v/G) - \log(M_s) + \\ \nonumber
 - \log(2\pi) - \log(I_e) - \log(R_e) = 0. 
\label{eqbet}
\end{eqnarray}

\noindent
Posing now: 

\begin{eqnarray}
A  & = & \log(I_e)+\log(G/k_v)+\log(M_s/L)+\log(2\pi)+ \\ \nonumber
   &   & \log(R_e)  \\ \nonumber
B  & = & - 2\log(2\pi) - 4\log(R_e)  \\ \nonumber
A' & = &  \log(\sigma)  \\ \nonumber
B' & = & 2\log(\sigma) - \log(G/k_v) - \log(M_s) - \log(2\pi) -  \\ \nonumber 
   &   & \log(I_e) - \log(R_e)  
   \label{eq3}
\end{eqnarray}
we obtain the following system:

\begin{eqnarray}
A\beta + 2\log(L'_0) + B = 0 \\
A'\beta + \log(L'_0) + B'= 0
\label{eqsyst}
\end{eqnarray}
\noindent 
with solutions:

\begin{eqnarray}
 \beta & = & \frac{-2\log(L'_0) - B}{A} \\ 
 \log(L'_0) & = &\frac{A'B/A - B'}{1-2A'/A}.
 \label{eq2}
\end{eqnarray}

The key result is that the  parameters $L$, $M_s$, \re, \Ie\ and $\sigma$ of a galaxy fully determine its evolution in FP as encoded in the parameters $\beta$ and $L'_0$.

\end{appendix}

\end{document}